# Host Stars and How Their Compositions Influence Exoplanets

Natalie R. Hinkel,[1,2] Allison Youngblood,[3] and Melinda Soares-Furtado[4,5,*]

[1]Physics & Astronomy Department, Louisiana State University, Baton Rouge, LA 70803, USA

[2]Southwest Research Institute, 6220 Culebra Rd, San Antonio, TX 78238, USA

[3]Exoplanets and Stellar Astrophysics Laboratory, NASA Goddard Space Flight Center, Greenbelt, MD 20771, USA

[4]Department of Astronomy, University of Wisconsin-Madison, 475 N. Charter St., Madison, WI 53703, USA

[5]Department of Physics and Kavli Institute for Astrophysics and Space Research, Massachusetts Institute of Technology, Cambridge, MA 02139, USA

## 1. INTRODUCTION

Though distant and seemingly unreachable, planets outside the Solar System, or exoplanets, have captivated the imagination of scientists and stargazers alike. With more than 5,000 confirmed exoplanet detections to date, it has become apparent that the Solar System — with multiple small, rocky planets interior to the larger gaseous planets — is not the only possible architecture for planetary systems. For example, some systems have "hot-Jupiters" where Jupiter-sized planets orbit very close to their host star (at distances comparable to the Sun-Mercury separation, e.g., Dawson & Johnson 2018). There are also planets that orbit two stars at the same time — much like Luke Skywalker's home planet Tatooine (e.g., Bromley & Kenyon 2015). Planets have been detected that are unlike anything in the Solar System, with sizes that are in between Earth and Neptune (known as super-Earths), meaning it is unclear whether they are giant rocky planets or small gaseous planets. In addition, because not all stars are like our yellow Sun, some planets orbit much redder stars (e.g., Dressing & Charbonneau 2013) or stars that barely shine at all because they are the remaining cores leftover from a stellar explosion (Wolszczan & Frail 1992). These diverse worlds serve as invaluable test cases to investigate their origin and the possibilities for planetary habitability. Stars are the building blocks of the Universe: they are the fundamental components of galaxies and the foundation of planetary systems. The size (mass and radius) of a star dictates its most basic properties such as temperature, evolution, and surface processes (like flares). All of these properties have a significant impact on an orbiting planet (Meadows & Barnes 2018). The composition of the star — the raw basic elements (like those found in the Periodic Table of Elements) — is linked to prior generations of stars that have seeded the cosmos with elements heavier than hydrogen and helium through their various nucleosynthesis[1] processes (e.g., Burbidge et al. 1957; Nomoto et al. 2013). Stars and their planets form at the same time and are made up of the same elements, gas, and dust. This chemical link between stars and planets is vital to determining the interior composition of small rocky planets. A planet's interior composition dictates whether properties that are essential to habitability are present, such as a clement climate, tectonics, and the presence of magnetic fields (Foley & Driscoll 2016). Unfortunately, these properties often cannot be directly measured with current technological capabilities. Therefore, the composition of the host star is used as a proxy for the make-up of the planet's interior.

Corresponding author: Natalie R. Hinkel natalie.hinkel@gmail.com

*NASA Hubble Postdoctoral Fellow

---

[1] Nucleosynthesis refers to the creation, or synthesis, of atomic nuclei.

It has become a common practice within the exoplanet field to say that "to know the star is to know the planet." The properties of the host star have a strong, direct influence on the interior and surface conditions of the orbiting planet and oftentimes measurements of planetary properties are made relative to the star's properties. Not only are observational measurements of the star necessary to determine even the most basic aspects of the planet (such as mass and radius), but the stellar environment influences how the planet evolves. Therefore, in this chapter, we begin by discussing the basics of stars, providing an overview of stellar formation, structure, photon and particle emissions, and evolution. Next, we go over the possible ways to determine the age of a star. We then outline how different kinds of stars are distributed within the Milky Way galaxy. Afterwards, we explain how to measure the composition of stars and the underlying math inherent to those observations, including caveats that are important when using the data for research applications. Finally, we explain the underlying physics and observations that enable stellar composition to be used as a proxy for planetary composition. In addition, given that this chapter focuses more on astronomy/astrophysics and uses a variety of important terms that may not be familiar to all readers, we have defined many terms either within the text or as a footnote for better interdisciplinary comprehension.

## 2. BASICS OF STARS

Stars are bright, shining balls of gas that are balanced under hydrostatic equilibrium, meaning the outward pressure created from their internal nuclear fusion is balanced by the inward pull of gravity. Note, however, that there are multiple phases throughout a star's lifetime where they are not fusing hydrogen into helium (Section 2.4) but are still called stars. Astronomers use a variety of parameters and classification schemes to describe stars, including spectral type, luminosity class, and the magnitude system[2]. Briefly, stars are grouped into spectral types (Capital letters: O, B, A, F, G, K, M) and luminosity classes (Roman numerals: I, II, III, IV, V, VI) based on the appearance of their optical or infrared spectra[3] (see the *Spectroscopy and Stellar Abundances* section). O-type stars are the largest, most massive, hottest, and shortest-lived stars, while M-type stars are the smallest, least massive, coolest, and longest-lived stars. The luminosity classes correspond to different evolutionary stages of a star. For instance, "V" corresponds to dwarf stars like our Sun that are in the longest-duration phase of their nuclear-burning lifetimes, akin to adulthood. Evolved stars are often called giants and have smaller Roman numerals. The Sun is a G V type dwarf star and the vast majority of known exoplanet host stars are F, G, K, and M-type stars of luminosity class V. In this section, we briefly describe the formation, structure, emissions and winds, and evolution of stars.

### 2.1. Formation

Stars are born in star-forming regions, where high-density molecular clouds[4] fragment into smaller regions of even higher density, ultimately forming many protostellar cores (e.g., Shu et al. 1987; McKee

---

[2] The magnitude system is a logarithmic measurement of stellar brightness where larger numbers correspond to fainter stars. Apparent magnitude describes how bright an object appears to us here on Earth and depends on the star's absolute magnitude (related to its intrinsic brightness), distance from the Earth, and the amount of interstellar dust between Earth and the object.

[3] A spectrum is a way to visualize a star's brightness as a function of wavelength and contains a wealth of information about a star, including its gravity, temperature, and elemental composition.

[4] Molecular clouds are cold, dense regions of gas and dust that provide the raw materials for star formation.

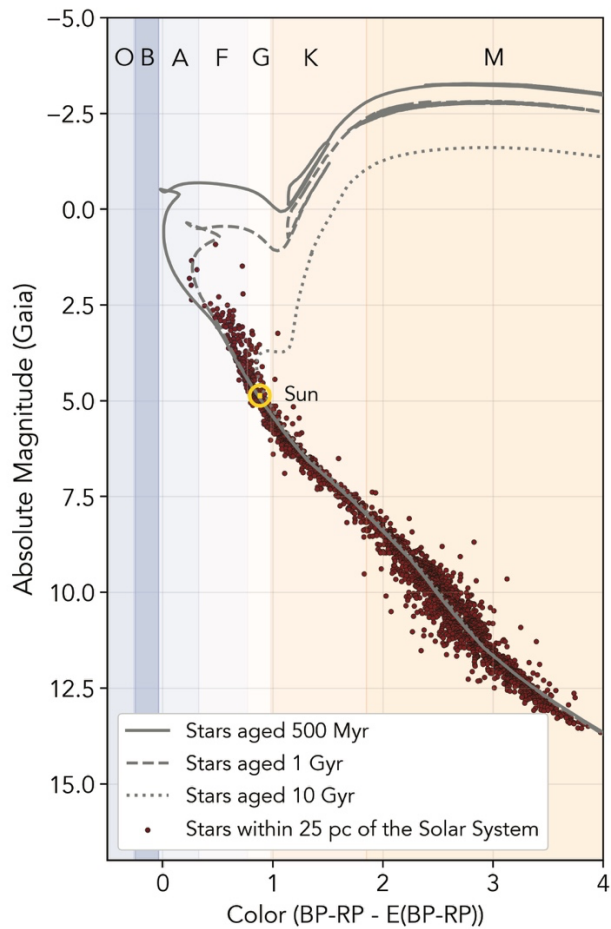

**Figure 1**. Color-magnitude diagram of stars within 25pc (1 parsec = $3.086 \times 10^{13}$ km) of our Solar System (Gaia Collaboration et al. 2016). The variable "BP-RP-E(BP-RP)" refers to the reddening-corrected color of a star, as measured by the Gaia space observatory (Gaia Collaboration et al. 2021). The location of the Sun is shown as a gold circle with a central dot (the symbol for the Sun). Like the Sun, the vast majority of stars in the solar neighborhood are on the main sequence. Also shown are the theoretically-determined locations of stars in a stellar system that is aged 500Myr (solid grey line), 1Gyr (dashed grey line), and 10Gyr (dotted grey line) (Bressan et al. 2012). The optical color of each of the spectral types is illustrated with visually representative colors (Harre & Heller 2021).

& Ostriker 2007; Luhman 2012). As these protostellar cores collapse under their own gravity, the conservation of angular momentum (or spin) leads to an increase in rotation rate and a protostellar disk forms with a protostar at the center. Gas from the protostellar nebula[5] is accreted[6] onto the protostar via the protostellar disk; however, the amount of material that can be accreted is oftentimes limited by powerful outflows from the protostar that emerge under conservation of angular momentum (e.g., Bally 2016; Hartmann et al. 2016).

As the protostar accretes more and more material, its internal temperature and pressure build until it reaches ~10 million K, which is the approximate threshold for the nuclear fusion process that turns hydrogen into helium (see, e.g., Burbidge et al. 1957). The protostar is now a fully formed star, termed a main sequence star for its position on the Hertzsprung-Russell (HR) diagram, which is a scatter plot of stellar luminosity and effective temperature. Figure 1 shows a color-magnitude diagram (CMD), which is similar to an observational HR diagram as color[7] is an observational proxy for effective temperature and absolute magnitude is directly related to luminosity. The timescale for stars to reach the main sequence ranges from less than 1 million years for the most massive stars to 1 billion years for the least massive stars. Note that young stars can deplete the gas in their protoplanetary disks (via accretion, winds, radiation pressure, and/or by planet formation) within only a few million years, while the planet formation process is thought to be complete within of order 1-10 million years (see, e.g., Lissauer 1993). The lifetime of protoplanetary disks may depend on stellar type, with more massive stars exhibiting shorter disk lifetimes (Ribas et al. 2015). Note also that more massive stars likely have more massive disks and therefore a greater reservoir of raw materials for planet formation (Andrews et al. 2013). This may explain why gas giants are rare around M-dwarfs, the lowest mass stars (e.g., Laughlin et al. 2004; Pass et al. 2023). However, lower mass stars are more likely to have at least one planet than higher mass stars (e.g., Mulders et al. 2015; Yang et al. 2020). Host star metallicity (or the composition of those elements heavier than H or He) also plays a role in the planet formation process, because stars with higher metallicity are more likely to have giant planets, presumably because their protoplanetary disks have more condensable solids (e.g., Gonzalez 1997; Fischer & Valenti 2005; Mulders et al. 2016). For an overview of the connection between planet formation theory and observed exoplanet demographic trends, see Gaudi et al. (e.g., 2021), Rodríguez Martínez et al. (e.g., 2023), and references within.

---

[5] A nebula is a region of densely clumped gas and dust often present before and after star formation.

[6] Accretion is the accumulation of matter, often under the influence of gravity, leading to growth of an object.

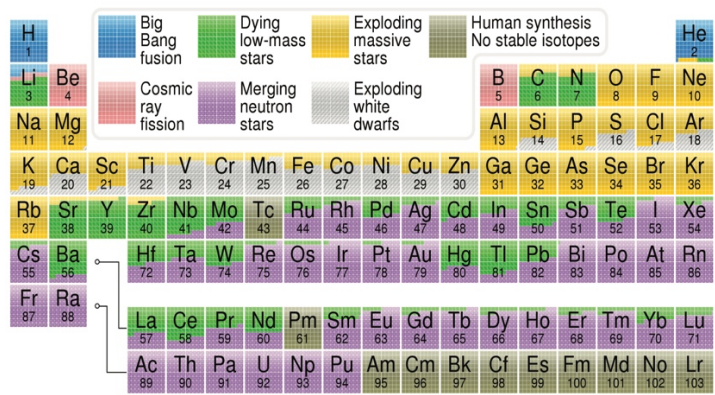

Figure 2. Periodic Table of the Elements color-coded to depict the primary formation channels that give rise to a given element. Note that this is a first-order simplification of a research field with many open questions, as well as a rough approximation of the data collected by Jennifer A. Johnson. While beryllium (Be) was produced in trace amounts during the Big Bang, the vast majority of Be is produced via cosmic ray fission (the splitting of an atomic nucleus by high-energy cosmic rays). Image credit: Jennifer Johnson, OSU.

### 2.2. Structure

Stars are roughly spherical objects comprised of ionized gas and are bound together by their own gravity. They are composed mostly of hydrogen and helium (99.9% by number or 98% by mass for the Sun), with a small but important percentage of metals. Astronomers refer to most elements heavier than helium as a metal because, aside from trace amounts of lithium and beryllium, they were not produced during the Big Bang but rather arise from nucleosynthesis in stellar cores, supernovae (powerful explosion of massive stars), or other processes (Burbidge et al. 1957; Nomoto et al. 2013; Arcones & Thielemann 2023; also see Fig. 2).

---

[7] An object's color is defined as the difference in apparent magnitude between two distinct bandpasses.

| Spectral Type | Effective Temp. (K) | Mass ($M_\odot$) | Main Sequence (MS) Lifetime | Fraction of MS Stars | End Product |
|---|---|---|---|---|---|
| O | >31,500 K | > 18 | <10Myr | 0.00003% | neutron star or black hole |
| B | $10,000 - 31,500$ | $2.7 - 18$ | $10 - 500$Myr | 0.13% | neutron star, black hole, or white dwarf |
| A | $7,500 - 10,000$ | $1.7 - 2.7$ | $500$Myr $- 2$Gyr | 0.6% | white dwarf |
| F | $6,000 - 7,500$ | $1.1 - 1.7$ | $2 - 7$Gyr | 3% | white dwarf |
| G | $5,300 - 6,000$ | $0.9 - 1.1$ | $7 - 15$Gyr | 7.6% | white dwarf |
| K | $3,900 - 5,300$ | $0.6 - 0.9$ | $15 - 60$Gyr | 12% | white dwarf |
| M | $2,300 - 3,700$ | $0.08 - 0.6$ | $> 60$Gyr | 76.5% | white dwarf |

Table 1. Properties corresponding to main sequence (MS) stars of a given spectral type (Pecaut & Mamajek 2013; Choi et al. 2016). Note that the most massive stars are thought to be limited to ∼150-300$M_\odot$ (Figer 2005; Crowther et al. 2010) based primarily on observations of the most massive known stellar clusters, and the least massive stars are limited to ∼0.08$M_\odot$ (Chabrier et al. 2000) because of the core temperature threshold requirement for fusion of hydrogen into helium. Objects below 0.08$M_\odot$ are called brown dwarfs, and their energy source is the fusion of deuterium into helium. Brown dwarfs are formed slightly less frequently than M-dwarfs, still making them extremely numerous throughout the galaxy.

Stars inherit the heavy elemental composition (or metallicity) and angular momentum properties of their star-forming region (although pockets of inhomogeneous or stochastic processes can alter these stellar properties). The structure of stars are separated into interiors and atmosphere, with a thin "surface" layer called a photosphere dividing the two (Fig. 3). A star's radius is defined from the core to the photosphere because it is the layer of the star that can be "seen" in visible light because photons can freely travel into space from the photosphere.

At the center of a star's interior is its core, where temperatures are ∼10 million K or more and hydrogen fuses into helium; fusion is the star's energy source. Nuclear fusion reactions are highly dependent on temperature and vary depending on the stellar type (e.g., Nomoto et al. 2013; Choi et al. 2016). More massive stars with hotter cores quickly burn through their vast hydrogen supplies. The smallest stars (M-dwarfs) have a total mass between approximately 0.08 and 0.50$M_\odot$ (in units of solar masses) but will take longer than the current age of the Universe to burn through their hydrogen supply and evolve past the main sequence. In other words, no M-dwarf has died! On the other hand, a B-type star with 10$M_\odot$ will burn through its hydrogen in only about 20Myr, which is very fast, astronomically-speaking. For reference, Table 1 lists the properties corresponding to the various spectral type classes[8] (effective temperatures, masses, main sequence lifetimes), as well as the fraction of stars in the Universe represented by a given spectral type, and the eventual end product after the star has ceased all nuclear-burning processes (e.g., Heger et al. 2003).

---

[8] https://www.pas.rochester.edu/~emamajek/EEM_dwarf_UBVIJHK_colors_Teff.txt

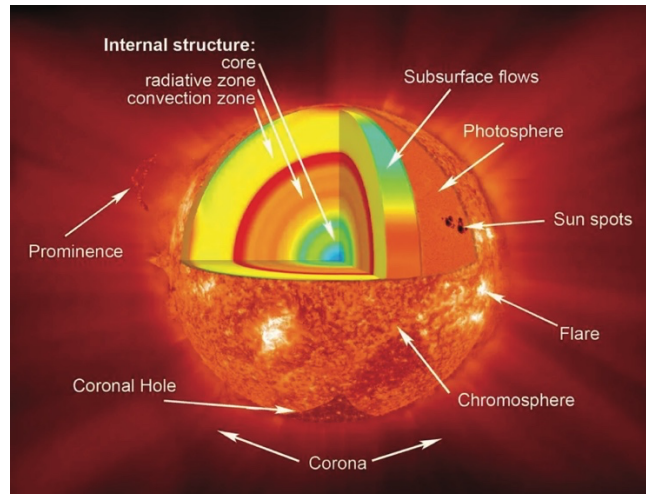

**Figure 3.** Schematic diagram of the Sun showing the interior and atmosphere. Image Credit: NASA.

Exterior to the stellar core is an envelope that carries the energy released from nuclear fusion to the photosphere via radiation (photons), convection (bulk motion of matter or plasma where material from deeper layers are moved vertically upwards into cooler layers and vice versa), or a mix of both (i.e., a radiation layer and/or a convection layer). The exact structure depends on the stellar mass (see Chabrier & Baraffe 1997; Paxton et al. 2011, 2018). The stellar temperature and density decline from the core to the photosphere. The temperature of the photosphere is given a special term called the effective temperature and is calculated through the Stefan-Boltzmann law, which describes the relationship between a star's luminosity (L; units: erg s$^{-1}$), radius (R; units: cm), and effective temperature (T$_{eff}$; units: K): L = $\sigma_{SB}$T$_{eff}^4$4$\pi$R$^2$, where $\sigma_{SB}$ is the Stefan-Boltzmann constant ($\sigma_{SB}$ = 5.67×10$^{-5}$ erg cm$^{-2}$ s$^{-1}$ K$^{-4}$). Between M-type and O-type stars, effective temperatures range from approximately 2000K to >30,000K (Table 1), respectively. For reference, the Sun's effective temperature is ∼5780K.

For the remainder of this section, we focus on the structure of the F, G, K, and M-type stars that comprise most of the known exoplanet host population. Above the photosphere is the stellar atmosphere, where the temperature continues to decline to a minimum before, surprisingly, rising again. The upper layers that comprise the stellar atmosphere in order of increasing altitude are called the chromosphere, transition region, and corona. The chromosphere contains the temperature minimum (Linsky 1980, 2017), and the corona contains the temperature maximum (Güdel & Nazé 2009); the transition region is aptly named as it is a narrow range of altitudes in the atmosphere where temperatures transition from chromospheric temperatures (∼10,000K) to coronal (∼1-10 million K). Note that stars more massive than ∼1.8M$_\odot$ (O, B, and some A types) are thought not to have chromospheres or coronae. The large temperatures of the upper atmosphere, despite the low density and great distance from the star's energy source, cannot be explained by radiative or convective heat transport from the core, but appear to be related to magnetism. F, G, K, and M-type stars generate large-scale magnetic fields; the magnetic field is generated in the hot, rotating stellar core, and the convective layers in the envelope transport and magnify these fields to the surface (Ossendrijver 2003). The energy carried by these magnetic fields is dissipated in the chromosphere through the corona; the exact nature of this dissipation and energy propagation is an active area of research (e.g., Zweibel & Yamada 2009).

## 2.3. Photon and Particle Emissions

The vast majority of a star's light emanates from its hot photosphere, which emits roughly as a blackbody (i.e., according to Planck's radiation law that defines the behavior of an object that absorbs all radiation and re-radiates that energy) with different atomic and molecular absorption features[9] (see the *Spectroscopy and Stellar Abundances* section) superimposed. Broadly speaking, the stellar metallicity does not dictate the most prominent absorption features, but rather the temperature and density do by controlling the dominant ionization states[10] or molecular species of the gas (see the *Stellar Abundance Caveats* section). As described by Wien's displacement law, a star's photospheric emission peaks in different parts of the electromagnetic spectrum depending on the stellar effective temperature, with the hottest stars emitting most of their photospheric radiation in the ultraviolet and the coolest stars emitting most of their energy in the infrared (see Fig. 1 for the colors of different spectral types as they would appear to the human eye). Cooler stellar photospheres are not warm enough to emit light in the ultraviolet, and no photospheres are hot enough to generate X-ray emission. Yet, many stars are UV and X-ray bright thanks to their magnetically-heated upper atmosphere (the chromosphere, transition region, and corona). Emission lines from atoms and ions tend to dominate the energy output in the UV and X-ray (Linsky 2017; Güdel & Nazé 2009). Sun-like stars (astronomers often refer to F, G, and K-type dwarfs as "Sun-like") emit ~10% of their total luminosity in the X-ray and UV, and M-dwarfs emit ~1% in the X-ray and UV. Although these emissions are small compared to the stellar luminosity, these UV and X-ray photons have a significant impact on the atmospheres and surface conditions of orbiting planets.

Stars also emit charged particles via winds and explosive eruptions. Massive stars' winds are driven by their intense radiation pressure (Castor et al. 1975), while low-mass stars like the Sun and M-dwarfs have hot, low-density outer layers (the corona) that expand, creating an outflow (Parker 1965). These stellar winds are streams of charged particles flowing away from the star through interplanetary space. Low-mass stars like the Sun also experience eruptions called coronal mass ejections, which are pockets of coronal material (ions) with their own magnetic fields that stream out into interplanetary space where they can interact with planets and their magnetic fields (Green et al. 2018). For an overview of winds and CMEs from exoplanet host stars, see Wood et al. (2021) and references therein.

## 2.4. Evolution

Stars are not static, homogeneous objects. They exhibit brightness variations over all time scales and have bright and dark regions spotting their surfaces (e.g., Penza et al. 2022). With timescales that range from seconds-to-hours, they exhibit brightness variations due to stellar flares (Benz & Güdel 2010) and acoustic standing waves excited by convection just beneath the stellar photosphere (see the *Age-Dating Stars* section and Kurtz 2022 for an overview of asteroseismology[11]). On days-to-months timescales, their brightness variations are due to the emergence, fading, or rotation (in and out of the line of sight of the observer) of starspots (dark spots) or faculae (bright spots). On multi-year timescales, stellar luminosity varies due to a more systematic change in the emergence of spots and faculae. For example, the Sun exhibits an 11-year magnetic activity cycle (Hathaway 2015); some active K and M-dwarfs have been found to have much shorter cycles, on the order of a few years (e.g., Boro Saikia et al. 2018).

---

[9] Absorption features show up as dark lines or dips in the spectrum when viewed using a spectrograph, which is a tool used to separate light into its different wavelengths.

[10] The process of gaining or losing electrons is known as ionization. There are different levels of ionization (ionization states) that an atom can undergo, resulting in the formation of ions with varying numbers of electrons.

[11] Asteroseismology is the study of the internal structure and properties of stars through the observation and analysis of their natural oscillations or vibrations.

Over the course of a star's long main-sequence lifetime (see Table 1 for typical values), stars will continue to rotate and fuse hydrogen into helium in their cores. As more of the material in the stellar core is converted into helium, fusion rates increase and stars become brighter over their lifetimes. For example, the Sun is 30% brighter today than it was 4.6 Gyr ago (Güdel 2007). Stellar winds can remove significant angular momentum thanks to the coupling between the star's magnetic field and the ionized material of the stellar wind and coronal mass ejections. The interaction of these magnetic fields generates a torque that slows down the rotation of the star over time; this well studied phenomenon is known as magnetic braking (Skumanich 1972). Eventually, when no more hydrogen remains for fusion, the star leaves the main sequence phase of its life and enters into stages marked by relatively rapid changes, which are highly dependent on the star's mass. In short, after passing through a red giant phase, stars that are less massive than $\sim 8 M_\odot$ will become compact stellar remnants, known as white dwarfs. Stars more massive than this approximate threshold will explode as supernovae and become neutron stars[12] or black holes. Note that white dwarfs in close binaries with red giants can accrete mass directly from the giant, exceeding a stability threshold and resulting in a supernova. Nucleosynthesis during these late stages of a star's life are important sources of many metals in the Universe. Remarkably, the first exoplanet system ever discovered was found orbiting a type of neutron star known as a pulsar (Wolszczan & Frail 1992). Exoplanets have also been observed around other late-stage hosts, such as red giants (Huber et al. 2013) and white dwarfs (Vanderburg et al. 2020). However, these exotic worlds represent a small minority of the known exoplanet population. There are active searches to detect more planets orbiting these late-stage hosts in hopes of building census demographics to help constrain theories of their formation and evolution.

## 3. AGE-DATING STARS

Determining both an accurate and precise age of a given star or stellar group is of critical significance in astronomy. These measurements have far-reaching implications for a wide range of phenomena in the Universe, setting constraints on the formation and evolution of planets, stars, clusters, and galaxies. There are a number of age-dating techniques available to the modern-day astronomer. These age estimates are generally determined using model-dependent or empirical methods; however, not all techniques are applicable to a given star. For an in-depth review of stellar age-dating, we refer the interested reader to Soderblom (2010, 2015).

Stars in controlled environments, such as stellar associations, co-moving groups, and clusters are among the most well-suited to age-dating analyses. Since these stars form together from a shared molecular cloud, they share a roughly common metallicity and are coeval (or of the same age, give or take a few million years). Stellar evolutionary models can be leveraged to age-date stars in such environments. This method relies on fitting stellar parameters within a CMD using model isochrones [13] (see Fig. 1 for tracks corresponding to three distinct isochronal ages; Soderblom 2010). This method is most applicable to coeval stellar populations of a known metallicity that exhibit a clearly discernible main-sequence turnoff[14] point. Knowing the metallicity of these environments is critical to accurate age-dating estimates using stellar evolutionary models, as a star's location on the CMD is metallicity-dependent. More specifically, stars that are metal-poor tend to appear brighter and bluer than their metal-rich counterparts, as light emanates from their photospheres more efficiently (due to reduced absorption and scattering).

---

[12] A neutron star is the dense remaining core leftover from a stellar supernova, see Table 1.

[13] Isochrones illustrate the CMD position of stars of a particular age, born from the same initial composition of elements.

CMD-determined age estimates are not well-suited for age-dating the most common star within our Milky Way galaxy: low-mass (K- and M-type stars) main sequence stars. This is because low-mass stars have long main-sequence lifetimes where their CMD positions do not significantly change over billions of years, making it hard to determine their specific age. So the CMD-position of many K- and M-type stars are consistent with stars (of a common mass) that span an age range of several billion years. As shown in Figure 1, there is an indistinguishable overlap in the CMD positions of low-mass main sequence stars corresponding to systems of considerably different ages (500Myr, 1Gyr, and 10Gyr).

While a star's CMD position does not change appreciably when it's on the main sequence, its rotation rate slows in a fairly predictable way, resulting in a gyrochronological[15] relation (color-rotation sequence) that provides a useful means of age-dating young ($\lesssim$ 1Gyr) low-mass (F, G, K-type) stars (Barnes 2003). This age-dating technique, which was first empirically identified by Skumanich (1972), leverages the relationship between a main sequence star's rotation period, its color (a useful proxy for mass, as shown in Fig. 1), and its age (the slope of the color-rotation relation changes with time). The gyrochronological relation corresponding to three clusters of differing ages is illustrated in Figure 4. Note, how the bluer, massive stars more readily converge onto a color-rotation relation — a result of more efficient angular momentum loss. The stellar rotation rates are often inferred from the photometric variability observed as starspots pass in and out of the line of sight of the observer, tracing out a regular rotation sequence. Such starspots are only present on stars with convective envelopes and reasonably rapid rotation rates (e.g., Berdyugina 2005). Therefore, this method is constrained to early-to-mid main sequence stars that have not lost considerable angular momentum ($\lesssim$ 1Gyr). While rotation age-dating is the most precise tool available among low-mass stars (K and M-type), there are also challenges in using this tool to age-date young ($\lesssim$ 100Myr), low-mass stars, as these stars take more time to shed angular momentum (via magnetic braking) and converge onto a well-behaved gyrochronological sequence (see Fig. 4).

Another useful age-dating technique relies on asteroseismic stellar data. More specifically, this method relies on variations in stellar brightness generated by oscillations on the surface of a star. The frequency spectrum of such photometric oscillations can be used to determine the mass, size, and internal properties (such as the depth of a convective zone) of a given target (Aerts et al. 2010). A star's observed oscillation frequencies are then compared to predicted frequencies (which vary with age-dependent changes in the star's size and internal structure), providing age estimates with a relatively high degree of accuracy (e.g., Soderblom 2010). Accurate asteroseismic age-dating requires high-quality observational data of a star's oscillation frequencies, which is generally not available for all stars of interest.

A star's spectroscopic element abundance signature can also offer a useful age indicator (see the *Spectroscopy and Stellar Abundances* section). More specifically, a main sequence star's lithium absorption strength is known to decrease with stellar age (e.g., Skumanich 1972; Carlos et al. 2016), likely due to extra mixing processes that remove lithium from the convective envelope of the star and/or depletion of this fragile trace element (e.g., Baraffe et al. 2017). Lithium is depleted via proton capture, but this process requires high temperatures (2.5 million Kelvin). As described in the *Structure* section, stellar core temperatures increase with stellar age, along with the temperature at the base of the convective envelope, as this region deepens with time (Iben 1967). The strength of a star's lithium feature is compared to another variable, such as stellar color, effective temperature, or rotation period, which is known to produce an age-dependent relation (e.g., Stanford-Moore et al. 2020). This method is optimal for low-mass ($< 1.5 M_\odot$), isolated, main sequence stars that do not exhibit anomalous rotation signatures. This is because the internal temperature, evolutionary state, rotational histories, and accretion histories impact the abundance signatures of lithium.

---

[14] The main sequence turnoff point is a key feature of a CMD, denoting the location where stars in a coeval stellar population begin to evolve away from the main sequence phase.

[15] Gyrochronology literally translates from Greek to mean "rotation age study."

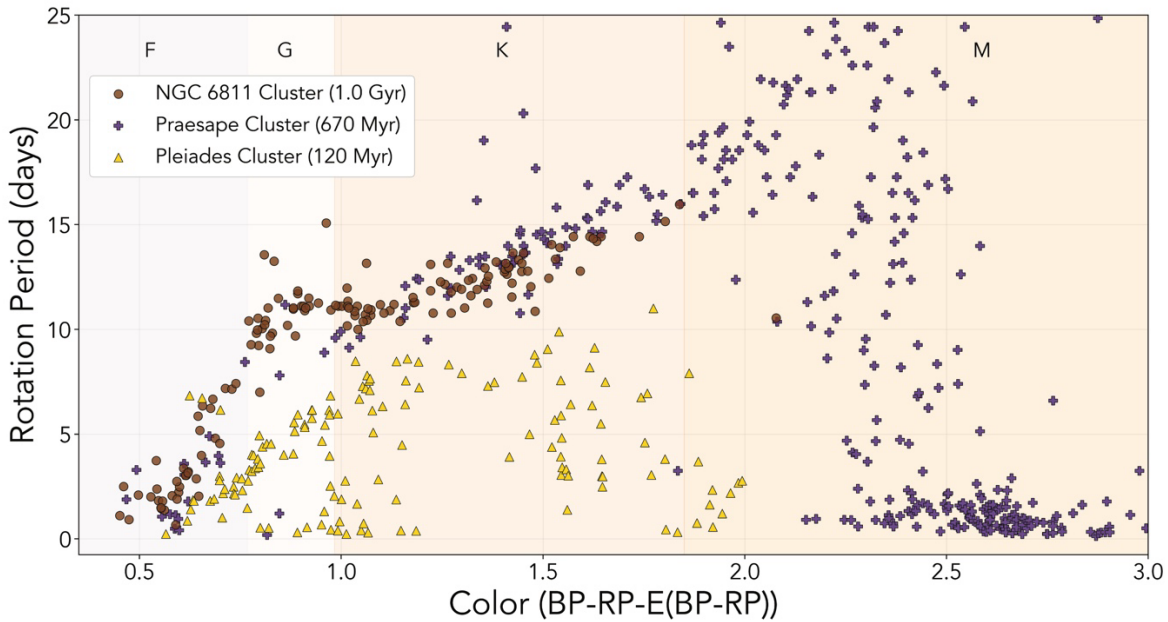

**Figure 4.** The gyrochronological rotation sequence of three known stellar clusters of differing ages: NGC 6811: 1Gyr (Curtis et al. 2019), Praesape: 670Myr (Douglas et al. 2017), and the Pleiades: 120Myr (Rebull et al. 2016). Clusters older than a few hundred Myr trace out a tighter rotation-color relation, as they have shed sufficient angular momentum for this pattern to emerge. Stars more massive than those depicted here generally do not exhibit the starspots responsible for photometric rotation measurements. Low-mass M-type stars take many hundreds of Myr to shed angular momentum via magnetic braking and, therefore, do not trace out the rotation sequence. The optical color of each of the spectral types is illustrated with visually representative hexadecimal colors (Harre & Heller 2021).

## 4. POPULATION DISTRIBUTION OF STARS IN THE GALAXY

There are approximately 100 billion stars in the Milky Way galaxy based on observed stellar densities and the structure of the galaxy. The vast majority of these stars formed in the Milky Way based on their kinematics (or the study of the position, velocity, and acceleration of objects over time) and/or metallicity, but interlopers have been observed and identified that have anomalous properties compared to other Milky Way stars.

Star-forming regions yield a regular distribution of stellar masses that appears to be independent of region-specific properties (e.g., size, metallicity), at least in the Milky Way (e.g., Bastian et al. 2010; Smith 2020). This distribution is referred to as the Initial Mass Function (Kroupa 2001; Chabrier 2003) and it accounts for the fact that a molecular cloud forms a small percentage of massive stars and an overwhelmingly large percentage of low-mass stars (see Table 1). More specifically, less than 4% of stars will be more massive than $1M_\odot$, while more than 75% will constitute stars lower than $0.5M_\odot$.

Many stars also have companion stars. The same gravitational fragmentation process that results in multiple protostars within a molecular cloud can result in gravitationally-bound protostars (Duchêne & Kraus 2013a). Or, massive protoplanetary disks can gravitationally fragment, producing companion stars or brown dwarfs. The stellar multiplicity[16] rate is not constant for all stars; it varies depending on stellar characteristics and environmental conditions (e.g., Duchêne & Kraus 2013b; Winters et al. 2019; Niu et al. 2021). The typical multiplicity rate for sun-like stars (in terms of mass) in our galaxy is ∼50% (e.g., Raghavan et al. 2010). Empirically it is observed that more massive stars tend to have companions at wider separations and less massive stars are more likely to have companions at closer separations (Duchêne & Kraus 2013a; Winters et al. 2019). Stars with companions are known to sometimes undergo mass exchange, mergers, and collisions, especially pre- or post-main sequence when their radii are substantially larger than their main sequence radii or when they reside in dense stellar environments where the frequency of a collision is higher. Mergers can result in nucleosynthesis, enriching future generations of stars with heavy elements. The existence of a companion star can have diverse effects on an exoplanet, influencing factors such as its dynamical stability, tidal heating, atmospheric conditions, and even the truncation of protoplanetary disks (Howell et al. 2022).

## 4.1. Abundance Gradient in the Milky Way

The first stars in the Universe formed out of the primordial gas leftover after the Big Bang (hydrogen, helium, and trace amounts of lithium and beryllium) and thus were extremely metal-poor (e.g., Bromm 2013). Without metals, stellar models predict that these first stars were ∼1000 times more massive than the Sun, had very short lives, and quickly became massive black holes (see the *Evolution* section above). These black holes gravitationally attracted nearby gas, which helped trigger the formation of new stars near the black hole — likely forming a central galactic bulge[17] as shown in Figure 5. Accreted gas persistently fed bulge star formation as well as the central black hole, increasing its gravitational influence. For the Milky Way, it is generally believed that large, independent clouds of matter and debris were caught up by the black hole's gravity over time and eventually created the extended galactic halo[18] (Fig. 5), triggering a short burst of star formation, producing low metallicity halo stars and globular clusters[19] (Freeman & Bland-Hawthorn 2002, and references therein).

At the same point in the Milky Way's history when the halo stars were formed, there was a large variation in the chemical composition (see *Spectroscopy and Stellar Abundances* section) of the halo stars, as compared to the bulge stars. For example, the various accretion events, gas dynamics, and overall star formation within the bulge resulted in rapid stellar formation and eventually post-main sequence evolution (death). The stellar death, often in the form of supernovae (Table 1), released new elements into the surrounding regions, thereby increasing the overall metallicity within the bulge.

---

[16] Stellar multiplicity refers to the phenomenon where two or more stars are gravitationally bound and orbit each other in a system.

[17] The bulge is a central, densely-populated region at the core of many spiral galaxies.

[18] The galactic halo is a large, spherical region surrounding spiral galaxies like the Milky Way. The halo contains older stars and globular clusters.

[19] A globular cluster is a gravitationally bound group of many thousands to millions of very old stars.

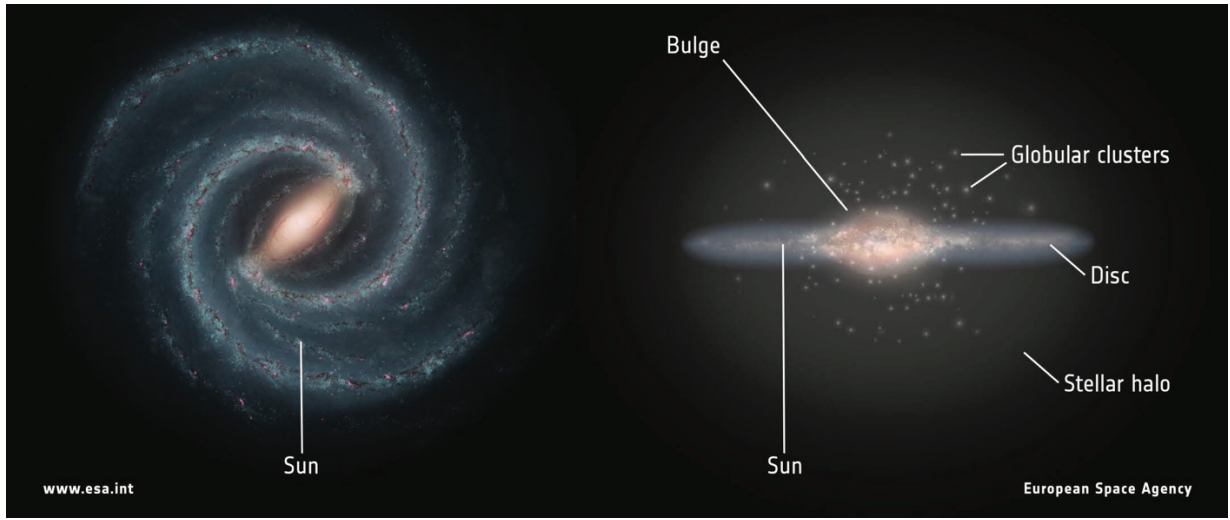

**Figure 5.** Diagram of the Milky Way spiral galaxy showing the key components such as the bulge, disk, and halo stars. Image credit left: NASA/JPL-Caltech; right: ESA; layout: ESA/ATG medialab.

The stars within the halo underwent a single, short (∼1-2 Gyr, Freeman & Bland-Hawthorn 2002) burst of star formation. Without the cycling of enriched material, the metallicity of the halo stars and globular clusters are more likely reflective of the smaller debris clouds that were pulled into the Galaxy's gravity well. In addition, they are notably the most metal-poor stars compared to all other stars within the galaxy — creating a metallicity or abundance gradient from the younger stellar bulge stars to the much older halo stars (Hawkins et al. 2015, and references therein).

After some time, and with the help of the rotation of the Milky Way, matter began to spread out into a plane to form the galactic disk (Fig. 5). Galactic disks in general are divided into two components, the thick disk and the thin disk — the latter is where the galactic bar and/or spiral arms[20] reside. In our galaxy, it is currently thought that the thick disk formed first, shortly after the settling of the disk (e.g., Masseron & Gilmore 2015). The thick disk is believed to have been formed from heating (e.g., causing an increase in temperature to) the early stellar disk via accretion events or minor mergers, for example with a low-mass galaxy (e.g., Walker et al. 1996; Rix & Bovy 2013), which created a vertical structure within the thick disk. Providing additional evidence that the thick disk stars are older, they have been empirically observed to have low iron-content (or [Fe/H], see the *Math Formalism of Stellar Abundances* section) and also being enriched in α-elements[21], such as C, O, Mg, Si, etc. (e.g., Rix & Bovy 2013; Bensby et al. 2014; Recio-Blanco et al. 2014; Hawkins et al. 2015; Ness et al. 2019; Bland-Hawthorn et al. 2019, and references therein).

In comparison, the thin disk is much flatter due to the higher rotational velocity around the Galactic Bulge. The majority of gas within the Milky Way resides in the thin disk, which resulted in many generations of star formation (i.e., the youngest stars) and stellar death, often via supernovae. Due to the many cycles of frequent star formation within the thin disk, the overall content of metals within the galaxy was increased over time (e.g., Bromm 2013). Another event crucial to the production of new elements within the thin disk is the merger of neutron stars. While rare, these merger events are impactful, leading to the ejection of neutron-rich material, which is subsequently mixed into the surrounding interstellar medium. In particular, supernovae and neutron star mergers led to the generation of rapid-neutron capture elements — elements that are formed by the rapid assembly of neutrons before the onset of radioactive decay. These elements are denoted in purple and yellow in the Periodic Table of Elements illustrated in Figure 2. Supernovae and neutron star merger events are responsible for producing most (stable) elements that are more massive than iron. As a result, thin disk stars (including the Sun) have the highest metallicity content compared to all other stars within the Galaxy.

---

[20] Galactic bars and arms are the build-up of dust and gas into thin regions in the galactic disk.

[21] Many of the elements with an even atomic number were formed by fusing multiple He atoms, known as the α-particle.

Especially from our vantage point on Earth, orbiting the Sun at 8.5kpc (1 parsec or pc = $3.086 \times 10^{13}$ km) from the center of the Milky Way (Fig. 5), we often rely on the observed kinematics of gas, dust, and stars to infer the overall make-up of our galaxy. And while there are many kinematic and chemical differences between the thin and thick disk stars, identifying the original birthplace of a star is not always obvious (e.g., Hawkins et al. 2015). For example, the vertical structure within the thick disk is at a maximum near the bulge and tapers with radial distance (e.g., Bensby et al. 2012; Bovy et al. 2012; Rix & Bovy 2013; Bovy et al. 2016), meaning that there are very few [Fe/H]-poor and α-rich stars in the outer disk of the Milky Way (>2kpc from the central bulge). By taking into account a stellar age gradient within the thick disk, Martig et al. (2016) were able to help reconcile the discrepancies within the definitions of the disk components. However, the Martig et al. (2016) result provides a strong implication that there are geometric, chemical, and age gradients throughout both disk components, making their distinction a sometimes difficult determination.

## 5. USING SPECTROSCOPY TO MEASURE THE COMPOSITION OF STARS

We have generally discussed the metallicity content of stars or their elemental composition, however, we have not yet gotten into the specifics. Therefore, in this section, we will go over how astronomers measure element abundances within stars via spectroscopy, the math underlying stellar abundances (e.g., the abundance of elements within stars) — including how to convert them to molar fractions, and caveats to fully understanding stellar abundance measurements.

### 5.1. Spectroscopy and Stellar Abundances

As mentioned above, stars are hot, dense balls of gas whose internal fusion reactions cause them to emit light as radiation. Measuring the intensity of the stellar light across a wavelength range, usually using an instrument called a spectrograph, results in a stellar spectrum. An object that is a perfect emitter of light or radiation, i.e., blackbody radiation, produces a continuous spectrum. However, stellar spectra are not continuous — they have dark spectral absorption lines due to the atoms (and sometimes molecules) that exist in the stellar photosphere. These absorption lines occur when the stellar radiation is absorbed by an atom, causing an electron to make the transition from a lower energy orbit to a higher orbit. Each atom absorbs light at unique wavelengths, depending on its atomic structure (electron shell configuration, quantum numbers, etc.), meaning that it's possible to identify which atoms are within the stellar photosphere via the absorption features. Because the presence of more atoms results in wider and deeper absorption features, it is possible to use spectroscopy to measure the area (or equivalent width) of the absorption features and determine the amount — or abundance — of an element within a star's photosphere.

In order to measure the spectroscopic abundances of elements within a star, first the fundamental properties of the star need to be understood — such as the effective temperature ($T_{eff}$), surface gravity (log g), and metallicity ([M/H], or [Fe/H] is often used as a proxy). These properties determine how radiation propagates and reacts within a star, ultimately defining the overall stellar atmospheric model. Jofré et al. (2019) — as part of a very thorough breakdown of current spectroscopic abundance techniques — provided a list of the current, regularly updated stellar models that are publicly available in their Table 1. In addition, the specifics of the atomic (and/or molecular) lines that are going to be measured need to be collected from theoretical calculations, published journal articles, or laboratory experiments. The actual stellar abundances can then be measured by using either 1) the "curve-of-growth" technique — which employs measuring the width of individual absorption lines, 2) a modeled synthetic spectrum that is compared to the measured spectrum, or 3) differential analysis to directly compare the differences in spectra between one star and another star on a line-by-line basis. Along with Jofré et al. (2019), we also refer the reader to Allende Prieto (2016) who provides an excellent explanation of the nuances in determining stellar abundances.

In addition to the differences between stellar abundance techniques, there are also variations between telescopes and their spectrographic instruments, which could have an impact on the measured abundances. For example, spectrograph resolution determines the extent to which spectral lines can be resolved, as well as the presence of lines that are blended together or lines that have asymmetries. This is of the utmost importance for precise abundance measurement (typically 0.05-0.2 dex, see the *Math Formalism of Stellar Abundances* section), meaning that most stellar abundances are measured using high-resolution instruments, which can have varying resolutions from R ($\Delta\lambda/\lambda$) = 50,000 - 100,000 in the optical band (while high resolution in the infrared is R $\gtrsim$ 20,000). Signal-to-noise (S/N) of a spectrograph defines the overall abundance precision, where a S/N > 100 is considered to be high.

Astronomers have been able to measure the spectroscopic (as opposed to photometric[22]) abundances of stars for decades. Over that time, a variety of stellar models, abundance techniques, spectrographs, and telescopes have been built. While these multitudes of methodologies provide an interesting way to observe, analyze, and compare stellar spectra, they also result in discrepancies. Fortunately, the community has come together in a variety of ways to better understand the differences and create important solutions. To begin, Smiljanic et al. (2014), Hinkel et al. (2016), and Jofré et al. (2017) studied a variety of abundance techniques within the community and analytically compared them to one another to better understand their strengths, weaknesses, inconsistencies, and overall relationships. At the same time, large spectroscopic surveys have been designed to obtain large, homogeneous datasets of thousands (if not millions) of stellar abundances. Some of the largest current or upcoming surveys are: RAdial Velocity Experiment (RAVE, R~7500) measuring 7 elements in ~300,000 stars (Steinmetz et al. 2006), WHT Enhanced Area Velocity Explorer (WEAVE, R ~ 20,000) will obtain abundances (the exact elements have not yet been specified) for ~1.5 million bright stars in the galactic field and in open clusters (Jin et al. 2022), the GALactic Archeology with HERMES (GALAH) survey will observe ~30 element abundances in an estimated ~1 million stars (currently at ~350,000 observed stars Buder et al. 2021), Apache Point Observatory Galaxy Evolution Experiment (APOGEE, R~22,500) has currently determined ~30 elements in ~350,000 stars (Buder et al.

---

[22] Spectroscopy measures the flux or intensity of light from a star with respect to wavelength, while photometry measures flux with respect to an area/region of space at a very narrow wavelength range, similar to a photograph.

2021), and the Gaia-ESO survey had a resolution of R ∼ 47,000 to observe ∼30 elements in ∼115,000 stars (Gilmore et al. 2012).

While large surveys succeed in creating consistent abundance data from a single source telescope+instrument combination, they often sacrifice element diversity and abundance precision – measuring a smaller number of elements while incurring larger measurement uncertainties — compared to smaller, more specific ground-based studies that often have higher resolution and S/N. To this end, the Hypatia Catalog (www.hypatiacatalog.com) is an amalgamate database compiled from ∼300 literature sources that measured high-resolution abundances for Fe in addition to one other element for main sequence (F-, G-, K-, M-type) stars within 500pc, or any exoplanet host regardless of distance (Hinkel et al. 2014). The Hypatia Catalog is the largest element abundance database for nearby stars, currently containing abundance measurements for ∼90 elements and species within ∼11,100 stars, ∼1450 of which are confirmed exoplanet host stars. There is a plethora of data within the Hypatia Catalog, which is uniquely multidimensional because it incorporates multiple measurements of the same element within the same star from different literature sources. Therefore, a concerted effort was made to ensure the abundance data is more homogeneous by standardizing the solar normalization scale (see the *Math Formalism of Stellar Abundances* section and Hinkel et al. 2014, 2022). Regardless of the heterogeneity, the Hypatia Catalog is a powerful tool because of the unique, unrivaled combination of the breadth (number of stars) and depth (number of elements).

## 5.2. Math Formalism of Stellar Abundances

Stellar abundances provide key insight into the chemical history and make-up of a stellar system, while also creating a compositional connection between stars and planets (see the *Composition of Stars and Their Planets* section). Therefore, it makes sense to understand how stellar abundances are mathematically defined, so that the measurements may be converted across disciplines. Here we provide a summary of the math formalism defined in Hinkel et al. (2022), which provides more detail as well as a walk-through example.

Stellar abundances are defined as a ratio between a generic element Q with respect to hydrogen (H), where the number of hydrogen atoms are considered to be a constant $10^{12}$ (Payne 1925a,b; Claas 1951), in a similar way that meteoritic abundances are often compared to $10^6$ silicon atoms. In this way, Q represents the number of Q atoms for every hydrogen atom – which is another way of showing that stellar abundances are ultimately the amount or number of an element (e.g., abundances are not to be confused with mass ratios). In terms of respective number fractions q and h — where a number fraction of 0 indicates that this element is not present and 1 means that an object is composed entirely of that element, we see that

$$Q' = q/h \times 10^{12}. \tag{1}$$

Now, we are able to define the absolute stellar abundance of Q, or A(Q), as

$$A(Q) \equiv \log_{10}(Q') , \tag{2}$$

where ≡ indicates that the two values are mathematically equivalent to one another. Looking at hydrogen in particular, we see that the absolute value A(H) = 12 and the number fraction h ≈ 1, since it is the most ubiquitous element in the Universe. This means that Eq. 2 is able to be defined as

$$\begin{aligned} A(Q) &= \log_{10}(q \times 10^{12}) \\ &= \log_{10}(q) + 12 \\ &= \log_{10}(q) + A(H) \\ &\equiv Q/H. \end{aligned} \tag{3}$$

Stellar abundances tend to be either defined with respect to H or Fe. Recognizing that Q/H = A(Q) and Fe/H = A(Fe), it is possible to convert between the two ratios using:

$$\begin{aligned} Q/Fe &\equiv Q/H - Fe/H \\ &= A(Q) - A(Fe). \end{aligned} \tag{4}$$

Because the Sun is the closest star, and therefore the most highly observed star, it is customary in astronomy to normalize stellar abundances with respect to the Sun. In this way, [Q/H] = 0.0dex means that a star has the same elemental abundance for Q as the Sun. Fortunately, normalizing to the Sun is a fairly intuitive subtraction in log-space:

$$\begin{aligned} [Q/H] &\equiv Q_*/H_* - Q_\odot/H_\odot \\ &= A(Q)_* - A(Q)_\odot, \end{aligned} \tag{5}$$

where the observed stellar abundances are indicated with an $*$, and the solar abundances are demarcated using $\odot$. We note here that logged abundances that are normalized to the Sun are defined with the "dex" unit, meaning "decadic logarithmic unit" (Lodders 2019). The "decadic" aspect of dex means that the log base is 10, similar to the decibel (dB), which is able to span a large dynamic range more easily defined using logarithmic (as opposed to linear) scales. See Figure 6 for an example plot showing stellar abundances for two ratios ([Si/Fe] vs [Fe/H]) in dex notation.

While working in the dex-notation for stellar abundances is useful for astronomers, it is often more appropriate for other fields to work with mole or molar ratios. It is, therefore, necessary to convert from log- to linear-space while also removing the implicit solar normalization. We convert from the original stellar abundance, [Q/H]$_*$, to moles, Q$_*$, using:

$$Q_* = 10^{([Q/H]_* + Q_\odot)}, \tag{6}$$

where Q$_\odot$ is the solar value used to normalize Q. We refer the reader to Hinkel et al. (2022) for an analysis of error propagation when converting from dex to moles (their Section 4.1) as well as a more in-depth discussion on the variety and comparison of solar normalizations within the stellar abundance field (their Section 6).

### 5.3. Stellar Abundance Caveats

When working with stellar abundance data, there are a number of technical considerations related to the underlying methods and models that need to be taken into account. To begin, most stellar models assume that small regions of the star's interior are in a thermally isolated, consistent (or steady-state) condition that can be well defined with equations, known as local thermodynamic equilibrium (LTE). However, a lot of physical processes happen within the stellar interior, creating varying spectral line strengths and unusual line shapes (Lodders 2019), indicating that there has been a deviation from LTE; these are called non-LTE (NLTE) effects. Introducing additional physics into the stellar models or investigating NLTE impacts on atomic and line data are time intensive, but also very important for resolving discrepancies between the data and models. Given the different underlying physical models between the calculations, we do not recommend combining (such as averaging) measurements that come from separate LTE and NLTE determinations. Their results, while comparable, are not equivalent.

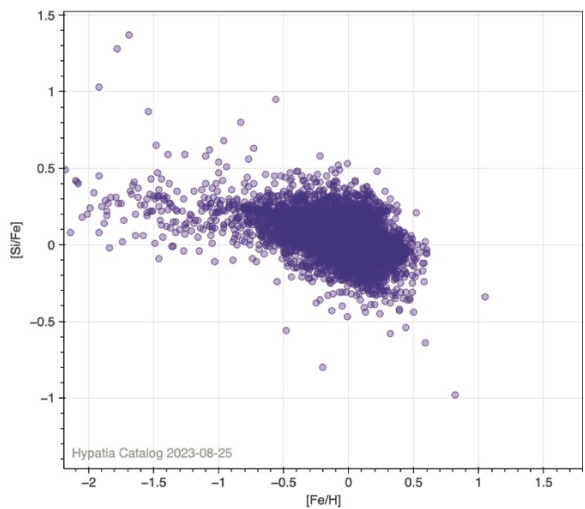

**Figure 6.** Classic stellar abundance plot (from the Hypatia Catalog, Hinkel et al. 2014) showing the ratio of [Si/Fe] with respect to [Fe/H], both with units in dex, where the Sun is located at (0, 0). Given the log-scale, a star with [Fe/H] = 0.5 dex has $10^{0.5} \approx 3.16$ times more Fe than the Sun.

In general, stellar abundances are defined as "elemental" abundances, meaning that they represent the overall amount of a specific element. While atomic lines are often used to determine abundances (mostly out of convenience), the term "elemental" doesn't preclude using molecular lines, such as FeH and $H_2O$, where the elements are disassociated to produce individual abundances, e.g., for Fe and O, respectively. The underlying implication is that, unless otherwise specified, the most common stable isotopes are usually measured and may be combined with other isotopes (Lodders 2019). Work is currently being done, for example by Lugaro et al. (2023), on separating stellar abundances into their distinct isotopic representations for certain elements.

Measuring different ionization states of an element — or when it's either gained an electron (anion) or lost an electron (cation) — can provide useful information for stellar abundance analysis. For example, many stellar abundance models iterate over the main stellar parameters ($T_{eff}$, log g, and [Fe/H]) until Fe I and Fe II are balanced (Jofré et al. 2015; Hinkel et al. 2016; Jofré et al. 2019). We note that astronomy ionization notation differs from chemistry such that a neutral element is indicated with a Roman numeral "I" (e.g, [Q I/H]) and a singly ionized element is denoted with "II" (e.g., [Q II/H]), an unnecessarily confusing alteration that is rooted in historical precedent and is therefore unlikely to be changed (Millikan & Bowen 1924). While some spectroscopists measure the ionization states separately, many combine them in the final measurement, often shown as [Q/H].

## 6. THE COMPOSITION OF STARS AND THEIR PLANETS

Nearly all elements in the Periodic Table were created in stars. Elements like Th and U are produced through rapid-process neutron capture and have been critical to the Earth's evolution. More specifically, the radioactive decay of Th and U (as well as $^{26}$Al and $^{60}$Fe) play a major role in heating the Earth's interior which is critical to mantle convection, the generation of the Earth's magnetic field, and the movement of the tectonic plates. Some rapid-process elements are vital to biological life, for example K is critical in maintaining proper cell function, nerve transmission, and muscle contraction. Stars within the Milky Way, even stars that are (currently) neighbors within the thin disk[23], have varying compositions that are the result of differing events that occurred prior to or during formation. As a result, the violent, seemingly rare astronomical events like supernovae and neutron star mergers seeded the Earth with elements that give rise to the everyday phenomena that make our lives possible.

### 6.1. Stellar Abundances as a Proxy for Planetary Interiors

It is not currently possible to directly measure the bulk mineralogical surface composition of an exoplanet, especially one with an atmosphere (Kreidberg et al. 2019; Zieba et al. 2023). While it's the goal for current and upcoming exoplanet missions like the James Webb Space Telescope (or JWST, launched 2021) and the Nancy Grace Roman Space Telescope (or Roman, expected launch 2026) to directly image planets and glean some sense of the planet's surface via color maps or measuring light reflected off the planet's surface (Lisse et al. 2020), it will be very difficult to obtain specific high-resolution data of the surface composition. To this end, no missions are currently expected to directly observe and measure a planet's interior composition and mineralogy.

Therefore, one of the most important tools currently available to understand the composition of a small planet is the relationship between its mass and radius, which helps define the planet's density and overall classification (e.g., super-Mercury, rocky, or mini-Neptune). While there are a variety of ways to detect an exoplanet, there are two that are the most common/popular: the radial-velocity technique that determines the gravitational wobble that the planet imparts on the star or the transit technique that measures how much stellar light is blocked when the planet passes in front of the star (from the Earth's perspective). Unfortunately, most current observing techniques for exoplanets make it very difficult to measure both planetary mass (radial-velocity) and radius (transit); for example, to date, only ∼20% of discovered planets within the NASA Exoplanet Archive[24] have both planet and mass measurements. In addition, planetary interior models based only on mass and radius are fraught with degeneracies. A study by Schlichting & Young (2022) illustrated the impact of elements that can be evaporated at low temperatures (or volatiles) on bulk densities, where hydrogen and other light elements within iron-rich cores significantly changed the mass-radius relationship for small planets. One way to break the mass-radius degeneracy is to employ the use of stellar abundances, especially those important to forming rocky material, as proxies for the composition of small planets (e.g., Lodders 2003; Sotin et al. 2007; Santos et al. 2015, 2017; Dorn et al. 2015, 2017a,b; Hinkel & Unterborn 2018; Putirka & Rarick 2019; Plotnykov & Valencia 2020; Putirka et al. 2021; Putirka & Xu 2021; Schulze et al. 2021).

The interiors of stars have high enough temperatures and pressures to support atomic fusion. And while some time-dependent processes slowly change the abundances of old stars, the composition of the stellar photosphere is largely reflective of the original molecular cloud from which it formed (Lodders et al. 2009; Lodders 2019). The stellar birth cloud also gave rise to other objects within a stellar system, such as orbiting planets, moons, asteroids, etc. Because they originate from the same place, it is possible to use the abundance of the host star as a proxy for the composition of their planets. This is especially true for elements that can only be evaporated at high temperatures (or refractory elements), such as Mg, Si, and Fe, which are not strongly impacted by chemical and physical processes within the stellar disk. For example, the molar ratios of these three elements within the Sun, Earth, and Mars are all the same to within 10% (McDonough 2003; Unterborn & Panero 2019). Importantly, Mg, Si, and Fe — in combination with O — are also especially important to the formation of small planets, making up 95mol % of the Earth (McDonough 2003).

---

[23] See abundances within the Hypatia Catalog (hypatiacatalog.com) to visualize the abundances differences between nearby stars.

[24] https://exoplanetarchive.ipac.caltech.edu/

From a theoretical perspective, the one-to-one relationship between stellar abundances and planetary interiors was studied by Bond et al. (2010a,b) who simulated the dynamics and chemistry when small planets are formed. Taking into account the changes within the stellar disk as the solar nebula evolved, they found that the bulk refractory abundances in small planets did not significantly change. In fact, their simulations lead them to believe that the differences in the interior composition of small exoplanets were more likely the result of variations between host star abundances than from processing within the disk. Thiabaud et al. (2015) specifically analyzed where and how the direct star-planet compositional link breaks down, modeling the Fe/Si, Mg/Si, and C/O ratios within the protoplanetary disk to see how it impacted the formation of rocky planets, ice planets, and giant gaseous planets. Comparing the composition of the three planet types to the abundances of the host star, they found that the Fe/Si and Mg/Si ratios matched between the star and planets.

More observationally, Bonsor et al. (2021) utilized an extremely interesting testbed involving a binary system, or two stars orbiting each other. One star is a normal main-sequence star while the other is a white dwarf star, which is the end-of-life result for stars that were originally $\sim$1-8M$_\odot$ (see Table 1); they are extremely dense, compact objects whose stellar spectra are fairly devoid of elemental or molecular lines. This particular white dwarf's spectrum has more metal absorption lines than a white dwarf should; the "pollution" indicates that a planetary object was accreted onto the surface of the white dwarf. Measuring the elemental abundances of polluted white dwarfs is the only direct way to determine the interior composition of the accreted object (e.g., Jura & Young 2014; Xu et al. 2017). However, chemical processing occurs during object accretion, which makes it difficult to identify the true composition of the original planetary body. Fortunately, the white dwarf (and planetary body) is in a binary orbit with a main sequence star, which all formed at the same time and should therefore exhibit similar abundances (e.g., Hawkins et al. 2020). This makes it possible to directly compare the composition of the planetary object to the original stellar composition. Through their analysis, Bonsor et al. (2021) found that the refractory abundances of the main-sequence Kdwarf star and the make-up of a comet-like body that had accreted onto the polluted white dwarf matched within measurement uncertainty. Guimond et al. (2023) expanded on these results by calculating the interior structures and mantle mineralogies of hypothetical planets using the host's stellar abundances. They found that not only did the bulk refractory compositions of stars and planets agree, but that the compositional similarity extended also to the mantle mineralogies of the exoplanets.

## 6.2. Additional Considerations for the Star-Planet Chemical Link

While stellar abundances are useful for a more holistic characterization and classification of an orbiting exoplanet, there are additional processes that occur during planet formation and evolution that alter the composition of a planet that must be taken into account. For example, Bonomo et al. (2019) analyzed two twin planets, Kepler-107 b and c, and found that the density of the c-planet was twice that of the b-planet.

After considering multiple possible reasons, it was found that the only likely mechanism to explain the difference was a giant impact that must have stripped much or all of the c-planet's silicate mantle. This is similar to the current understanding of why Mercury has a dissimilar Si, Mg, and Fe abundance compared to other Solar System bodies and is markedly iron-rich compared with the Earth and Mars (Benz et al. 2008).

In addition to the possibility of giant impacts, the temperature of the stellar disk controls the condensation of elements, which creates radial variations in different locations of the disk that change over time (e.g., Lodders 2003). While these radial variations are most likely to impact volatile elements, as opposed to refractory, Dorn et al. (2019) found that a possible exception occurs when refractory elements partially condense at temperatures > 1200K within the inner part of the stellar disk. They found that it is within this regime that super-Earths, such as HD 219134 b, 55 Cnc e, and WASP-47 e, could have potentially formed at high condensation temperatures (T > 1200 K) which results in Fe-depleted cores and enrichments in Al and Ca. Especially for super-Earth-type planets, which have no Solar System analog, it may be that the interior composition models are not better constrained by the inclusion of stellar abundances. This finding was supported by Otegi et al. (2020) who specifically studied stellar Fe/Si and Mg/Si molar ratios for planets with masses < 25 M$_\oplus$ and radii < 3.5 R$_\oplus$, although their results were dependent on uncertainties and the specific stellar abundance values.

Similarly, Adibekyan et al. (2021a,b) found that a radial gradient that is oxidizing within the disk, or possible Fe-enrichment coupled with a giant impact, may create differences during small planet formation. They used stoichiometry to balance mass fraction equations for those species dominant in a small planet, e.g., H, He, C, O, Mg, Si, and Fe (Santos et al. 2015). By analyzing the composition of 38 small planets, they tested the iron mass fraction and found that relationship between stellar abundances and planet composition was 4-to-1 instead of 1-to-1.

### 6.2.1. Free-Floating Planets

There is also the case of free-floating planets (FFPs): planetary-mass objects that do not orbit a star. Several hundred FFPs have been detected in young star-forming regions, such as ρ Ophiuchus (Miret-Roig et al. 2022; Bouy et al. 2022), Upper Scorpius (David et al. 2019; Lodieu et al. 2021; Miret-Roig et al. 2022; Bouy et al. 2022), Orion Trapezium Cluster (Lucas & Roche 2000), σ Orionis (Peña Ramírez et al. 2012), the Carina-Near moving group (Gagné et al. 2017), and NGC 1333 (Scholz et al. 2012). The age constraints provided by extremely young star-forming environments in which FFPs are found (commonly ranging between 1-10Myr), indicate that the FFP formation process must occur rapidly. In fact, a recent investigation of the Trapezium Cluster (1-3Myr) using the James Webb Space Telescope Near Infrared Camera Pearson & McCaughrean 2023 reported a population of 540 FFPs down to 0.6M$_J$ within an 11′ × 7.5′ (1.2pc × 0.8pc) field of view.

Like all planets, the internal structure and composition of FFPs are tightly linked to the origin and evolution of these bodies. Some FFPs are expected to form via typical stellar disk accretion processes, only to be ejected from their host stars due to dynamical processes (e.g., stellar fly-bys, disk instabilities, or planet scattering events, Rasio & Ford 1996; Veras & Raymond 2012). Like their bound counterparts, gas giants formed in this manner can widely range in metallicity, depending on the initial disk location and the properties of the host star (e.g., Öberg et al. 2011; Helling et al. 2014; Mordasini et al. 2016). Some FFPs may have never been bound to a central star, forming via the disk fragmentation of a molecular cloud (similar to stars) and would contribute to the low-mass end of the Initial Mass Function. Like ejected planets, FFPs formed via instabilities in the disk are predicted to exhibit a wide range in composition.

However, unlike ejected planets, they would share similar kinematic properties with co-moving neighbors fashioned from the same molecular cloud.

FFPs are valuable in ways that extend beyond their origin story. For example, they can serve as critical testbeds to probe the atmospheric properties of planets spared from the radiative effects of a central star, which are well known to be a dominant environmental influence. Contemporary and forthcoming surveys including JWST and Roman are expected to provide the FFP census data required to better investigate the origin of starless worlds.

In addition, the number of FFPs are certainly plentiful. Extrapolating their detections, an estimated several billion FFPs are expected to roam throughout the Milky Way galaxy (Miret-Roig et al. 2022). In fact, tens of thousands of FFPs have been predicted to occupy a spherical volume centered on Earth and extending to our nearest stellar neighbor, Proxima Centauri (d = 1.3pc; Lingam et al. 2023). JWST is the first instrument capable of directly detecting and characterizing FFPs, probing masses in the $1\text{-}15 M_J$ mass regime. As such, several JWST observing programs have focused on an FFP search within nearby star-forming regions. Similarly, forthcoming microlensing observations conducted by the Roman are expected to significantly increase the number of detected FFPs, probing masses down to $0.01 M_\oplus$ (e.g., Dai & Guerras 2018; Limbach et al. 2023).

Being able to characterize and classify small planets accurately — ones that orbit their host star and those that are free-floating — is of the utmost importance to upcoming NASA and ESA missions. It currently appears that the general, first-order assumption that stellar abundances can be used as a proxy for planet composition is valid until it (obviously) is not, e.g., for super-Earths and gas giants. However, it is vital to the community that we better understand crucial second- and third-order processes that occur within the disk and during planet formation itself (see Zhang & Trapman 2024, this volume; Mordasini & Burn 2024, this volume), since they have an influential impact on the planet's make-up.

## ACKNOWLEDGMENTS

The research shown here acknowledges the use of the Hypatia Catalog Database, an online compilation of stellar abundance data as described in Hinkel et al. (2014, AJ, 148, 54), which was supported by NASA's Nexus for Exoplanet System Science (NExSS) research coordination network and the Vanderbilt Initiative in Data-Intensive Astrophysics (VIDA). MSF gratefully acknowledges the generous support provided by NASA through Hubble Fellowship grant HST-HF2-51493.001-A awarded by the Space Telescope Science Institute, which is operated by the Association of Universities for Research in Astronomy, In., for NASA, under the contract NAS 5-26555.

## REFERENCES

Adibekyan V, Santos NC, Dorn C, et al. (2021a) Composition of super-Earths, super-Mercuries, and their host stars. Communications of the Byurakan Astrophysical Observatory 68:447-453, doi:10.52526/25792776-2021.68.2-44710.48550/arXiv.2112.14512

Adibekyan V, Dorn C, Sousa SG, et al. (2021b) A compositional link between rocky exoplanets and their host stars. Science 374:330-332, doi:10.1126/science.abg879410.48550/arXiv.2102.12444

Aerts C, Christensen-Dalsgaard J, Kurtz DW (2010) Asteroseismology.

Allende Prieto C (2016) Solar and stellar photospheric abundances. Living Reviews in Solar Physics 13:1, doi:10.1007/s41116-016-0001-6

Andrews SM, Rosenfeld KA, Kraus AL, Wilner DJ (2013) The Mass Dependence between Protoplanetary Disks and their Stellar Hosts. Astrophysical Journal 771:129, doi:10.1088/0004-637X/771/2/129

Arcones A, Thielemann F-K (2023) Origin of the elements. Astronomy & Astrophysicsr 31:1, doi:10.1007/s00159-022-00146-x

Bally J (2016) Protostellar Outflows. Annual Review of Astronomy and Astrophysics 54:491-528, doi:10.1146/annurev-astro-081915-023341

Baraffe I, Pratt J, Goffrey T, Constantino T, Folini D, Popov MV, Walder R, Viallet M (2017) Lithium Depletion in Solar-like Stars: Effect of Overshooting Based on Realistic Multi-dimensional Simulations. Astrophysical Journal Letters 845:L6, doi:10.3847/2041-8213/aa82ff

Barnes SA (2003) On the Rotational Evolution of Solar- and Late-Type Stars, Its Magnetic Origins, and the Possibility of Stellar Gyrochronology. Astrophysical Journal 586:464-479, doi:10.1086/367639

Bastian N, Covey KR, Meyer MR (2010) A Universal Stellar Initial Mass Function? A Critical Look at Variations. Annual Review of Astronomy and Astrophysics 48:339-389, doi:10.1146/annurev-astro-082708-101642

Bensby T, Feltzing S, Oey MS (2014) Exploring the Milky Way stellar disk. A detailed elemental abundance study of 714 F and G dwarf stars in the solar neighbourhood. Astronomy & Astrophysics 562:A71, doi:10.1051/0004-6361/201322631

Bensby T, Alves-Brito A, Oey MS, Yong D, Meléndez J (2012) Abundance Trends in the Inner and Outer Galactic Disk. In Book Abundance Trends in the Inner and Outer Galactic Disk. Vol 458. Editor, p 171

Benz AO, Güdel M (2010) Physical Processes in Magnetically Driven Flares on the Sun, Stars, and Young Stellar Objects. Annual Review of Astronomy and Astrophysics 48:241-287, doi:10.1146/annurev-astro-082708-101757

Benz W, Anic A, Horner J, Whitby JA (2008) The Origin of Mercury. In: Mercury. Vol 26. Balogh A, Ksanfomality L, von Steiger R, (eds). Space Science Reviews, p 7

Berdyugina SV (2005) Starspots: A Key to the Stellar Dynamo. Living Reviews in Solar Physics 2:8, doi:10.12942/lrsp-2005-8

Bland-Hawthorn J, Sharma S, Tepper-Garcia T, et al. (2019) The GALAH survey and Gaia DR2: dissecting the stellar disc's phase space by age, action, chemistry, and location. Monthly Notices of the Royal Astronomical Society 486:1167-1191, doi:10.1093/mnras/stz217

Bond JC, Lauretta DS, O'Brien DP (2010a) Making the Earth: Combining dynamics and chemistry in the Solar System. Icarus 205:321-337, doi:10.1016/j.icarus.2009.07.037

Bond JC, O'Brien DP, Lauretta DS (2010b) The Compositional Diversity of Extrasolar Terrestrial Planets. I. In Situ Simulations. Astrophysical Journal 715:1050-1070, doi:10.1088/0004-637X/715/2/1050

Bonomo AS, Zeng L, Damasso M, et al. (2019) A giant impact as the likely origin of different twins in the Kepler-107 exoplanet system. Nature Astronomy 3:416-423, doi:10.1038/s41550-018-0684-9

Bonsor A, Jofré P, Shorttle O, Rogers LK, Xu S, Melis C (2021) Host-star and exoplanet compositions: a pilot study using a wide binary with a polluted white dwarf. Monthly Notices of the Royal Astronomical Society 503:1877-1883, doi:10.1093/mnras/stab370

Boro Saikia S, Marvin CJ, Jeffers SV, Reiners A, Cameron R, Marsden SC, Petit P, Warnecke J, Yadav AP (2018) Chromospheric activity catalogue of 4454 cool stars. Questioning the active branch of stellar activity cycles. Astronomy & Astrophysics 616:A108, doi:10.1051/0004-6361/201629518

Bouy H, Tamura M, Barrado D, et al. (2022) Infrared spectroscopy of free-floating planet candidates in Upper Scorpius and Ophiuchus. Astronomy & Astrophysics 664:A111, doi:10.1051/0004-6361/202243850

Bovy J, Rix H-W, Hogg DW (2012) The Milky Way Has No Distinct Thick Disk. Astrophysical Journal 751:131, doi:10.1088/0004-637X/751/2/131

Bovy J, Rix H-W, Schlafly EF, Nidever DL, Holtzman JA, Shetrone M, Beers TC (2016) The Stellar Population Structure of the Galactic Disk. Astrophysical Journal 823:30, doi:10.3847/0004-637X/823/1/30

Bressan A, Marigo P, Girardi L, Salasnich B, Dal Cero C, Rubele S, Nanni A (2012) PARSEC: stellar tracks and isochrones with the PAdova and TRieste Stellar Evolution Code. Monthly Notices of the Royal Astronomical Society 427:127-145, doi:10.1111/j.1365-2966.2012.21948.x

Bromley BC, Kenyon SJ (2015) Planet Formation around Binary Stars: Tatooine Made Easy. Astrophysical Journal 806:98, doi:10.1088/0004-637X/806/1/98

Bromm V (2013) Formation of the first stars. Reports on Progress in Physics 76:112901, doi:10.1088/0034-4885/76/11/112901

Buder S, Sharma S, Kos J, et al. (2021) The GALAH+ survey: Third data release. Monthly Notices of the Royal Astronomical Society 506:150-201, doi:10.1093/mnras/stab1242

Burbidge EM, Burbidge GR, Fowler WA, Hoyle F (1957) Synthesis of the Elements in Stars. Reviews of Modern Physics 29:547-650, doi:10.1103/RevModPhys.29.547

Carlos M, Nissen PE, Meléndez J (2016) Correlation between lithium abundances and ages of solar twin stars. Astronomy & Astrophysics 587:A100, doi:10.1051/0004-6361/201527478

Castor JI, Abbott DC, Klein RI (1975) Radiation-driven winds in Of stars. Astrophysical Journal 195:157-174, doi:10.1086/153315

Chabrier G (2003) Galactic Stellar and Substellar Initial Mass Function. Publications of the Astronomical Society of the Pacific 115:763-795, doi:10.1086/376392

Chabrier G, Baraffe I (1997) Structure and evolution of low-mass stars. Astronomy & Astrophysics 327:1039-1053, doi:10.48550/arXiv.astro-ph/9704118

Chabrier G, Baraffe I, Allard F, Hauschildt P (2000) Evolutionary Models for Very Low-Mass Stars and Brown Dwarfs with Dusty Atmospheres. Astrophysical Journal 542:464-472, doi:10.1086/309513

Choi J, Dotter A, Conroy C, Cantiello M, Paxton B, Johnson BD (2016) Mesa Isochrones and Stellar Tracks (MIST). I. Solar-scaled Models. Astrophysical Journal 823:102, doi:10.3847/0004-637X/823/2/102

Claas WJ (1951) The composition of the solar atmosphere. University of Utrecht, Netherlands

Crowther PA, Schnurr O, Hirschi R, Yusof N, Parker RJ, Goodwin SP, Kassim HA (2010) The R136 star cluster hosts several stars whose individual masses greatly exceed the accepted 150M_solar stellar mass limit. Monthly Notices of the Royal Astronomical Society 408:731-751, doi:10.1111/j.1365-2966.2010.17167.x

Curtis JL, Agüeros MA, Douglas ST, Meibom S (2019) A Temporary Epoch of Stalled Spin-down for Low-mass Stars: Insights from NGC 6811 with Gaia and Kepler. Astrophysical Journal 879:49, doi:10.3847/1538-4357/ab2393

Dai X, Guerras E (2018) Probing Extragalactic Planets Using Quasar Microlensing. Astrophysical Journal Letters 853:L27, doi:10.3847/2041-8213/aaa5fb

David TJ, Hillenbrand LA, Gillen E, Cody AM, Howell SB, Isaacson HT, Livingston JH (2019) Age Determination in Upper Scorpius with Eclipsing Binaries. Astrophysical Journal 872:161, doi:10.3847/1538-4357/aafe09

Dawson RI, Johnson JA (2018) Origins of Hot Jupiters. Annual Review of Astronomy and Astrophysics 56:175-221, doi:10.1146/annurev-astro-081817-051853

Dorn C, Hinkel NR, Venturini J (2017a) Bayesian analysis of interiors of HD 219134b, Kepler-10b, Kepler-93b, CoRoT-7b, 55 Cnc e, and HD 97658b using stellar abundance proxies. Astronomy & Astrophysics 597:A38, doi:10.1051/0004-6361/20162874910.48550/arXiv.1609.03909

Dorn C, Harrison JHD, Bonsor A, Hands TO (2019) A new class of Super-Earths formed from high-temperature condensates: HD219134 b, 55 Cnc e, WASP-47 e. Monthly Notices of the Royal Astronomical Society 484:712-727, doi:10.1093/mnras/sty343510.48550/arXiv.1812.07222

Dorn C, Khan A, Heng K, Connolly JAD, Alibert Y, Benz W, Tackley P (2015) Can we constrain the interior structure of rocky exoplanets from mass and radius measurements? Astronomy & Astrophysics 577:A83, doi:10.1051/0004-6361/201424915

Dorn C, Venturini J, Khan A, Heng K, Alibert Y, Helled R, Rivoldini A, Benz W (2017b) A generalized Bayesian inference method for constraining the interiors of super Earths and sub-Neptunes. Astronomy & Astrophysics 597:A37, doi:10.1051/0004-6361/20162870810.48550/arXiv.1609.03908

Douglas ST, Agüeros MA, Covey KR, Kraus A (2017) Poking the Beehive from Space: K2 Rotation Periods for Praesepe. Astrophysical Journal 842:83, doi:10.3847/1538-4357/aa6e52

Dressing CD, Charbonneau D (2013) The Occurrence Rate of Small Planets around Small Stars. Astrophysical Journal 767:95, doi:10.1088/0004-637X/767/1/95

Duchêne G, Kraus A (2013) Stellar Multiplicity. Annual Review of Astronomy and Astrophysics 51:269-310, doi:10.1146/annurev-astro-081710-102602

Figer DF (2005) An upper limit to the masses of stars. Nature 434:192-194, doi:10.1038/nature03293
Fischer DA, Valenti J (2005) The Planet-Metallicity Correlation. ApJ 622:1102--1117

Foley BJ, Driscoll PE (2016) Whole planet coupling between climate, mantle, and core: Implications for rocky planet evolution. Geochemistry, Geophysics, Geosystems 17:1885--1914

Freeman K, Bland-Hawthorn J (2002) The New Galaxy: Signatures of Its Formation. Annual Review of Astronomy and Astrophysics 40:487-537, doi:10.1146/annurev.astro.40.060401.093840

Gagné J, Faherty JK, Burgasser AJ, Artigau É, Bouchard S, Albert L, Lafrenière D, Doyon R, Bardalez

Gagliuffi DC (2017) SIMP J013656.5+093347 Is Likely a Planetary-mass Object in the Carina-Near Moving Group. Astrophysical Journal Letters 841:L1, doi:10.3847/2041-8213/aa70e2

Gaia Collaboration, Brown AGA, Vallenari A, et al. (2016) Gaia Data Release 1. Summary of the astrometric, photometric, and survey properties. Astronomy & Astrophysics 595:A2, doi:10.1051/0004-6361/201629512

Gaia Collaboration, Brown AGA, Vallenari A, et al. (2021) Gaia Early Data Release 3. Summary of the contents and survey properties. Astronomy & Astrophysics 649:A1, doi:10.1051/0004-6361/202039657

Gaudi BS, Meyer M, Christiansen J (2021) The Demographics of Exoplanets. In: ExoFrontiers; Big Questions in Exoplanetary Science. Madhusudhan N, (ed), p 2-1

Gilmore G, Randich S, Asplund M, et al. (2012) The Gaia-ESO Public Spectroscopic Survey. The Messenger 147:25-31

Gonzalez G (1997) The stellar metallicity-giant planet connection. Monthly Notices of the Royal Astronomical Society 285:403-412, doi:10.1093/mnras/285.2.403

Green LM, Török T, Vrvsnak B, Manchester W, Veronig A (2018) The Origin, Early Evolution and Predictability of Solar Eruptions. Space Science Reviews 214:46, doi:10.1007/s11214-017-0462-5

Güdel M (2007) The Sun in Time: Activity and Environment. Living Reviews in Solar Physics 4:3, doi:10.12942/lrsp-2007-3

Güdel M, Nazé Y (2009) X-ray spectroscopy of stars. Astronomy & Astrophysicsr 17:309-408, doi:10.1007/s00159-009-0022-4

Guimond CM, Shorttle O, Rudge JF (2023) Mantle mineralogy limits to rocky planet water inventories. Monthly Notices of the Royal Astronomical Society 521:2535-2552, doi:10.1093/mnras/stad148

Harre J-V, Heller R (2021) Digital color codes of stars. Astronomische Nachrichten 342:578-587, doi:10.1002/asna.202113868

Hartmann L, Herczeg G, Calvet N (2016) Accretion onto Pre-Main-Sequence Stars. Annual Review of Astronomy and Astrophysics 54:135-180, doi:10.1146/annurev-astro-081915-023347

Hathaway DH (2015) The Solar Cycle. Living Reviews in Solar Physics 12:4, doi:10.1007/lrsp-2015-4

Hawkins K, Jofré P, Masseron T, Gilmore G (2015) Using chemical tagging to redefine the interface of the Galactic disc and halo. Monthly Notices of the Royal Astronomical Society 453:758-774, doi:10.1093/mnras/stv1586

Hawkins K, Lucey M, Ting Y-S, Ji A, Katzberg D, Thompson M, El-Badry K, Teske J, Nelson T, Carrillo A (2020) Identical or fraternal twins? The chemical homogeneity of wide binaries from Gaia DR2. Monthly Notices of the Royal Astronomical Society 492:1164-1179, doi:10.1093/mnras/stz3132

Heger A, Fryer CL, Woosley SE, Langer N, Hartmann DH (2003) How Massive Single Stars End Their Life. Astrophysical Journal 591:288-300, doi:10.1086/375341

Helling C, Woitke P, Rimmer PB, Kamp I, Thi W-F, Meijerink R (2014) Disk Evolution, Element Abundances and Cloud Properties of Young Gas Giant Planets. Life 4:142-173, doi:10.3390/life4020142

Hinkel NR, Unterborn CT (2018) The Star-Planet Connection. I. Using Stellar Composition to Observationally Constrain Planetary Mineralogy for the 10 Closest Stars. Astrophysical Journal 853:83, doi:10.3847/1538-4357/aaa5b4

Hinkel NR, Young PA, Wheeler CH, III (2022) A Concise Treatise on Converting Stellar Mass Fractions to Abundances to Molar Ratios. Astronomical Journal 164:256, doi:10.3847/1538-3881/ac9bfa

Hinkel NR, Timmes FX, Young PA, Pagano MD, Turnbull MC (2014) Stellar Abundances in the Solar Neighborhood: The Hypatia Catalog. Astronomical Journal 148:54, doi:10.1088/0004-6256/148/3/54

Hinkel NR, Young PA, Pagano MD, et al. (2016) A Comparison of Stellar Elemental Abundance Techniques and Measurements. Astrophysical Journal Supplement 226:4, doi:10.3847/0067-0049/226/1/4

Howell SB, Matson RA, Marzari F (2022) Editorial: The Effect of Stellar Multiplicity on Exoplanetary Systems. Frontiers in Astronomy and Space Sciences 8:830980, doi:10.3389/fspas.2021.830980

Huber D, Carter JA, Barbieri M, et al. (2013) Stellar Spin-Orbit Misalignment in a Multiplanet System. Science 342:331-334, doi:10.1126/science.1242066

Iben I, Jr. (1967) Stellar Evolution.VI. Evolution from the Main Sequence to the Red-Giant Branch for Stars of Mass 1 M_Sun, 1.25 M_Sun, and 1.5 M_Sun. Astrophysical Journal 147:624, doi:10.1086/149040

Jin S, Trager SC, Dalton GB, et al. (2022) The wide-field, multiplexed, spectroscopic facility WEAVE: Survey design, overview, and simulated implementation. arXiv e-prints:arXiv:2212.03981

Jofré P, Heiter U, Soubiran C (2019) Accuracy and Precision of Industrial Stellar Abundances. Annual Review of Astronomy and Astrophysics 57:571-616, doi:10.1146/annurev-astro-091918-104509

Jofré P, Heiter U, Worley CC, et al. (2017) Gaia FGK benchmark stars: opening the black box of stellar element abundance determination. Astronomy & Astrophysics 601:A38, doi:10.1051/0004-6361/201629833

Jofré P, Heiter U, Soubiran C, et al. (2015) Gaia FGK benchmark stars: abundances of alpha and iron-peak elements. Astronomy & Astrophysics 582:A81, doi:10.1051/0004-6361/201526604

Jura M, Young ED (2014) Extrasolar Cosmochemistry. Annual Review of Earth and Planetary Sciences 42:45-67, doi:10.1146/annurev-earth-060313-054740

Kreidberg L, Koll DDB, Morley C, et al. (2019) Absence of a thick atmosphere on the terrestrial exoplanet LHS 3844b. Nature 573:87-90, doi:10.1038/s41586-019-1497-4

Kroupa P (2001) On the variation of the initial mass function. Monthly Notices of the Royal Astronomical Society 322:231-246, doi:10.1046/j.1365-8711.2001.04022.x

Kurtz DW (2022) Asteroseismology Across the Hertzsprung-Russell Diagram. Annual Review of Astronomy and Astrophysics 60:31-71, doi:10.1146/annurev-astro-052920-094232

Laughlin G, Bodenheimer P, Adams FC (2004) The Core Accretion Model Predicts Few Jovian-Mass Planets Orbiting Red Dwarfs. Astrophysical Journal Letters 612:L73-L76, doi:10.1086/424384

Limbach MA, Soares-Furtado M, Vanderburg A, et al. (2023) The TEMPO Survey. I. Predicting Yields of Transiting Exosatellites, Moons, and Planets from a 30 days Survey of Orion with the Roman Space Telescope. Publications of the Astronomical Society of the Pacific 135:014401, doi:10.1088/1538-3873/acafa4

Lingam M, Hein AM, Eubanks TM (2023) Chasing Nomadic Worlds: A New Class of Deep Space Missions. arXiv e-prints:arXiv:2307.12411, doi:10.48550/arXiv.2307.12411

Linsky JL (1980) Stellar chromospheres. Annual Review of Astronomy and Astrophysics 18:439-488, doi:10.1146/annurev.aa.18.090180.002255

Linsky JL (2017) Stellar Model Chromospheres and Spectroscopic Diagnostics. Annual Review of Astronomy and Astrophysics 55:159-211, doi:10.1146/annurev-astro-091916-055327

Lissauer JJ (1993) Planet formation. Annual Review of Astronomy and Astrophysics 31:129-174, doi:10.1146/annurev.aa.31.090193.001021

Lisse CM, Desch SJ, Unterborn CT, Kane SR, Young PR, Hartnett HE, Hinkel NR, Shim S-H, Mamajek EE, Izenberg NR (2020) A Geologically Robust Procedure for Observing Rocky Exoplanets to Ensure that Detection of Atmospheric Oxygen Is a Modern Earth-like Biosignature. Astrophysical Journal Letters 898:L17, doi:10.3847/2041-8213/ab9b91

Lodders K (2003) Solar System Abundances and Condensation Temperatures of the Elements. Astrophysical Journal 591:1220-1247, doi:10.1086/375492

Lodders K (2019) Solar Elemental Abundances. The Oxford Research Encyclopedia of Planetary Science, Oxford University Press:arXiv:1912.00844

Lodders K, Palme H, Gail H-P (2009) Abundances of the Elements in the Solar System. Landolt Boumlrnstein 4B:712, doi:10.1007/978-3-540-88055-4_34

Lodieu N, Hambly NC, Cross NJG (2021) Exploring the planetary-mass population in the Upper Scorpius association. Monthly Notices of the Royal Astronomical Society 503:2265-2279, doi:10.1093/mnras/stab401

Lucas PW, Roche PF (2000) A population of very young brown dwarfs and free-floating planets in Orion. Monthly Notices of the Royal Astronomical Society 314:858-864, doi:10.1046/j.1365-8711.2000.03515.x

Lugaro M, Ek M, Pet Ho M, Pignatari M, Makhatadze GV, Onyett IJ, Schönbächler M (2023) Representation of s-process abundances for comparison to data from bulk meteorites. European Physical Journal A 59:53, doi:10.1140/epja/s10050-023-00968-y

Luhman KL (2012) The Formation and Early Evolution of Low-Mass Stars and Brown Dwarfs. Annual Review of Astronomy and Astrophysics 50:65-106, doi:10.1146/annurev-astro-081811-125528

Martig M, Minchev I, Ness M, Fouesneau M, Rix H-W (2016) A Radial Age Gradient in the Geometrically Thick Disk of the Milky Way. Astrophysical Journal 831:139, doi:10.3847/0004-637X/831/2/139

Masseron T, Gilmore G (2015) Carbon, nitrogen and alpha-element abundances determine the formation sequence of the Galactic thick and thin discs. Monthly Notices of the Royal Astronomical Society 453:1855-1866, doi:10.1093/mnras/stv1731

McDonough WF (2003) 2.15 - Compositional Model for the Earth's Core. In: Treatise on Geochemistry. Pergamon, Oxford, p 547-568

McKee CF, Ostriker EC (2007) Theory of Star Formation. Annual Review of Astronomy and Astrophysics 45:565-687, doi:10.1146/annurev.astro.45.051806.110602

Meadows VS, Barnes RK (2018) Factors Affecting Exoplanet Habitability. In: Handbook of Exoplanets. Deeg HJ, Belmonte JA, (eds). SpringerLink, p 57

Millikan RA, Bowen IS (1924) Extreme Ultra-violet Spectra. Physical Review 23:1-34, doi:10.1103/PhysRev.23.1

Miret-Roig N, Bouy H, Raymond SN, et al. (2022) A rich population of free-floating planets in the Upper Scorpius young stellar association. Nature Astronomy 6:89-97, doi:10.1038/s41550-021-01513-x

Mordasini C, van Boekel R, Mollière P, Henning T, Benneke B (2016) The Imprint of Exoplanet Formation History on Observable Present-day Spectra of Hot Jupiters. Astrophysical Journal 832:41, doi:10.3847/0004-637X/832/1/41

Mulders GD, Pascucci I, Apai D (2015) A Stellar-mass-dependent Drop in Planet Occurrence Rates. Astrophysical Journal 798:112, doi:10.1088/0004-637X/798/2/112

Mulders GD, Pascucci I, Apai D, Frasca A, Molenda-Żakowicz J (2016) A Super-solar Metallicity for Stars with Hot Rocky Exoplanets. Astronomical Journal 152:187, doi:10.3847/0004-6256/152/6/187

Ness MK, Johnston KV, Blancato K, Rix H-W, Beane A, Bird JC, Hawkins K (2019) In the Galactic Disk, Stellar [Fe/H] and Age Predict Orbits and Precise [X/Fe]. Astrophysical Journal 883:177, doi:10.3847/1538-4357/ab3e3c

Niu Z, Yuan H, Wang S, Liu J (2021) Binary Fractions of G and K Dwarf Stars Based on Gaia EDR3 and LAMOST DR5: Impacts of the Chemical Abundances. Astrophysical Journal 922:211, doi:10.3847/1538-4357/ac2573

Nomoto Ki, Kobayashi C, Tominaga N (2013) Nucleosynthesis in Stars and the Chemical Enrichment of Galaxies. Annual Review of Astronomy and Astrophysics 51:457-509, doi:10.1146/annurev-astro-082812-140956

Öberg KI, Murray-Clay R, Bergin EA (2011) The Effects of Snowlines on C/O in Planetary Atmospheres. Astrophysical Journal Letters 743:L16, doi:10.1088/2041-8205/743/1/L16

Ossendrijver M (2003) The solar dynamo. Astronomy & Astrophysicsr 11:287-367, doi:10.1007/s00159-003-0019-3

Otegi JF, Dorn C, Helled R, Bouchy F, Haldemann J, Alibert Y (2020) Impact of the measured parameters of exoplanets on the inferred internal structure. Astronomy & Astrophysics 640:A135, doi:10.1051/0004-6361/20203800610.48550/arXiv.2006.12353

Parker EN (1965) The passage of energetic charged particles through interplanetary space. Planetary and Space Science 13:9-49, doi:10.1016/0032-0633(65)90131-5

Pass EK, Winters JG, Charbonneau D, Irwin JM, Latham DW, Berlind P, Calkins ML, Esquerdo GA, Mink J (2023) Mid-to-late M Dwarfs Lack Jupiter Analogs. Astronomical Journal 166:11, doi:10.3847/1538-3881/acd349

Paxton B, Bildsten L, Dotter A, Herwig F, Lesaffre P, Timmes F (2011) Modules for Experiments in Stellar Astrophysics (MESA). Astrophysical Journal Supplement 192:3, doi:10.1088/0067-0049/192/1/3

Paxton B, Schwab J, Bauer EB, et al. (2018) Modules for Experiments in Stellar Astrophysics (MESA): Convective Boundaries, Element Diffusion, and Massive Star Explosions. Astrophysical Journal Supplement 234:34, doi:10.3847/1538-4365/aaa5a8

Payne CH (1925a) Stellar Atmospheres; a Contribution to the Observational Study of High Temperature in the Reversing Layers of Stars. Radcliffe College.

Payne CH (1925b) Astrophysical Data Bearing on the Relative Abundance of the Elements. Proceedings of the National Academy of Science 11:192-198, doi:10.1073/pnas.11.3.192

Pearson SG, McCaughrean MJ (2023) Jupiter Mass Binary Objects in the Trapezium Cluster. arXiv e-prints:arXiv:2310.01231, doi:10.48550/arXiv.2310.01231

Pecaut MJ, Mamajek EE (2013) Intrinsic Colors, Temperatures, and Bolometric Corrections of Pre-main-sequence Stars. Astrophysical Journal Supplement 208:9, doi:10.1088/0067-0049/208/1/9

Peña Ramírez K, Béjar VJS, Zapatero Osorio MR, Petr-Gotzens MG, Martín EL (2012) New Isolated Planetary-mass Objects and the Stellar and Substellar Mass Function of the sigma Orionis Cluster. Astrophysical Journal 754:30, doi:10.1088/0004-637X/754/1/30

Penza V, Berrilli F, Bertello L, Cantoresi M, Criscuoli S, Giobbi P (2022) Total Solar Irradiance during the Last Five Centuries. Astrophysical Journal 937:84, doi:10.3847/1538-4357/ac8a4b

Plotnykov M, Valencia D (2020) Chemical fingerprints of formation in rocky super-Earths' data. Monthly Notices of the Royal Astronomical Society 499:932-947, doi:10.1093/mnras/staa261510.48550/arXiv.2010.06480

Putirka K, Xu S (2021) On the Lithology and Mineralogy of Polluted White Dwarf Materials. Bulletin of the American Astronomical Society 53:1044

Putirka KD, Rarick JC (2019) The composition and mineralogy of rocky exoplanets: A survey of >4000 stars from the Hypatia Catalog. American Mineralogist 104:817-829, doi:10.2138/am-2019-6787

Putirka KD, Dorn C, Hinkel NR, Unterborn CT (2021) Compositional Diversity of Rocky Exoplanets. Elements 17:235, doi:10.48550/arXiv.2108.08383

Raghavan D, McAlister HA, Henry TJ, Latham DW, Marcy GW, Mason BD, Gies DR, White RJ, ten Brummelaar TA (2010) A Survey of Stellar Families: Multiplicity of Solar-type Stars. Astrophysical Journal Supplement 190:1-42, doi:10.1088/0067-0049/190/1/1

Rasio FA, Ford EB (1996) Dynamical instabilities and the formation of extrasolar planetary systems. Science 274:954-956, doi:10.1126/science.274.5289.954

Rebull LM, Stauffer JR, Bouvier J, et al. (2016) Rotation in the Pleiades with K2. I. Data and First Results. Astronomical Journal 152:113, doi:10.3847/0004-6256/152/5/113

Recio-Blanco A, de Laverny P, Kordopatis G, et al. (2014) The Gaia-ESO Survey: the Galactic thick to thin disc transition. Astronomy & Astrophysics 567:A5, doi:10.1051/0004-6361/201322944

Ribas Á, Bouy H, Merín B (2015) Protoplanetary disk lifetimes vs. stellar mass and possible implications for giant planet populations. Astronomy & Astrophysics 576:A52, doi:10.1051/0004-6361/201424846

Rix H-W, Bovy J (2013) The Milky Way's stellar disk. Mapping and modeling the Galactic disk. Astronomy & Astrophysicsr 21:61, doi:10.1007/s00159-013-0061-8

Rodríguez Martínez R, Martin DV, Gaudi BS, Schulze JG, Asnodkar AP, Boley KM, Ballard S (2023) A Comparison of the Composition of Planets in Single-planet and Multiplanet Systems Orbiting M dwarfs. Astronomical Journal 166:137, doi:10.3847/1538-3881/aced9a

Santos NC, Adibekyan V, Dorn C, et al. (2017) Constraining planet structure and composition from stellar chemistry: trends in different stellar populations. Astronomy & Astrophysics 608:A94, doi:10.1051/0004-6361/20173135910.48550/arXiv.1711.00777

Santos NC, Adibekyan V, Mordasini C, et al. (2015) Constraining planet structure from stellar chemistry: the cases of CoRoT-7, Kepler-10, and Kepler-93. Astronomy & Astrophysics 580:L13, doi:10.1051/0004-6361/201526850

Schlichting HE, Young ED (2022) Chemical equilibrium between Cores, Mantles, and Atmospheres of Super-Earths and Sub-Neptunes, and Implications for their Compositions, Interiors and Evolution. Planetary Science Journal 9:19 pp., doi:10.48550/arxiv.2107.10405

Scholz A, Jayawardhana R, Muzic K, Geers V, Tamura M, Tanaka I (2012) Substellar Objects in Nearby Young Clusters (SONYC). VI. The Planetary-mass Domain of NGC 1333. Astrophysical Journal 756:24, doi:10.1088/0004-637X/756/1/24

Schulze JG, Wang J, Johnson JA, Gaudi BS, Unterborn CT, Panero WR (2021) On the Probability That a Rocky Planet's Composition Reflects Its Host Star. The Planetary Science Journal 2:113, doi:10.3847/PSJ/abcaa8

Shu FH, Adams FC, Lizano S (1987) Star formation in molecular clouds: observation and theory. Annual Review of Astronomy and Astrophysics 25:23-81, doi:10.1146/annurev.aa.25.090187.000323

Skumanich A (1972) Time Scales for Ca II Emission Decay, Rotational Braking, and Lithium Depletion. Astrophysical Journal 171:565, doi:10.1086/151310

Smiljanic R, Korn AJ, Bergemann M, et al. (2014) The Gaia-ESO Survey: The analysis of high-resolution UVES spectra of FGK-type stars. Astronomy & Astrophysics 570:A122, doi:10.1051/0004-6361/201423937

Smith RJ (2020) Evidence for Initial Mass Function Variation in Massive Early-Type Galaxies. Annual Review of Astronomy and Astrophysics 58:577-615, doi:10.1146/annurev-astro-032620-020217

Soderblom DR (2010) The Ages of Stars. Annual Review of Astronomy and Astrophysics 48:581-629, doi:10.1146/annurev-astro-081309-130806

Soderblom DR (2015) Ages of Stars: Methods and Uncertainties. In Book Ages of Stars: Methods and Uncertainties. Vol 39. Editor, p 3

Sotin C, Grasset O, Mocquet A (2007) Mass radius curve for extrasolar Earth-like planets and ocean planets. Icarus 191:337-351, doi:10.1016/j.icarus.2007.04.006

Stanford-Moore SA, Nielsen EL, De Rosa RJ, Macintosh B, Czekala I (2020) BAFFLES: Bayesian Ages for Field Lower-mass Stars. Astrophysical Journal 898:27, doi:10.3847/1538-4357/ab9a35

Steinmetz M, Zwitter T, Siebert A, et al. (2006) The Radial Velocity Experiment (RAVE): First Data Release. Astronomical Journal 132:1645-1668, doi:10.1086/506564

Thiabaud A, Marboeuf U, Alibert Y, Leya I, Mezger K (2015) Elemental ratios in stars vs planets. Astronomy & Astrophysics 580:A30, doi:10.1051/0004-6361/201525963

Unterborn CT, Panero WR (2019) The Pressure and Temperature Limits of Likely Rocky Exoplanets. Journal of Geophysical Research (Planets) 124:1704-1716, doi:10.1029/2018JE005844

Vanderburg A, Rappaport SA, Xu S, et al. (2020) A giant planet candidate transiting a white dwarf. Nature 585:363-367, doi:10.1038/s41586-020-2713-y

Veras D, Raymond SN (2012) Planet-planet scattering alone cannot explain the free-floating planet population. Monthly Notices of the Royal Astronomical Society 421:L117-L121, doi:10.1111/j.1745-3933.2012.01218.x

Walker IR, Mihos JC, Hernquist L (1996) Quantifying the Fragility of Galactic Disks in Minor Mergers. Astrophysical Journal 460:121, doi:10.1086/176956

Winters JG, Henry TJ, Jao W-C, Subasavage JP, Chatelain JP, Slatten K, Riedel AR, Silverstein ML, Payne MJ (2019) The Solar Neighborhood. XLV. The Stellar Multiplicity Rate of M Dwarfs Within 25 pc. Astronomical Journal 157:216, doi:10.3847/1538-3881/ab05dc

Wolszczan A, Frail DA (1992) A planetary system around the millisecond pulsar PSR1257 + 12. Nature 355:145-147, doi:10.1038/355145a0

Wood BE, Müller H-R, Redfield S, et al. (2021) New Observational Constraints on the Winds of M dwarf Stars. Astrophysical Journal 915:37, doi:10.3847/1538-4357/abfda5

Xu S, Zuckerman B, Dufour P, Young ED, Klein B, Jura M (2017) The Chemical Composition of an Extrasolar Kuiper-Belt-Object. Astrophysical Journal Letters 836:L7, doi:10.3847/2041-8213/836/1/L7

Yang J-Y, Xie J-W, Zhou J-L (2020) Occurrence and Architecture of Kepler Planetary Systems as Functions of Stellar Mass and Effective Temperature. Astronomical Journal 159:164, doi:10.3847/1538-3881/ab7373

Zieba S, Kreidberg L, Ducrot E, et al. (2023) No thick carbon dioxide atmosphere on the rocky exoplanet TRAPPIST-1 c. Nature 620:746-749, doi:10.1038/s41586-023-06232-z

Zweibel EG, Yamada M (2009) Magnetic Reconnection in Astrophysical and Laboratory Plasmas. Annual Review of Astronomy and Astrophysics 47:291-332, doi:10.1146/annurev-astro-082708-101726